%Paper: hep-ph/9305319
%From: Peter Landshoff <P.V.Landshoff@damtp.cambridge.ac.uk>
%Date: Wed, 26 May 93 14:10:22 BST
%Date (revised): Mon, 12 Jul 93 16:40:44 BST

\magnification\magstep1
\tolerance 10000

\font\rfont=cmr10 at 10 true pt
\def\ref#1{$^{\hbox{\rfont {[#1]}}}$}

  %%Fonts
\font\fourteenrm=cmr12 scaled\magstep1

\font\fourteenit=cmti12 scaled\magstep1

  %%Greek

\def\pmb#1{\setbox0=\hbox{#1}% \kern-.025em\copy0\kern-\wd0
 \kern.05em\copy0\kern-\wd0 \kern-.025em\raise.0433em\box0 }

  %%Fractions

 %

 %%FORMATTING
\def \i {\item} \def \ii {\itemitem}

\parskip=6pt
\parindent=0pt
\hsize=17truecm\hoffset=-5truemm
\voffset=-1truecm\vsize=24.5truecm
\def\footnoterule{\kern-3pt
\hrule width 17truecm \kern 2.6pt}

  %%REFERENCES
%     \defref\label{text}
% generates a number, assigns it to \label, generates an entry.
% To list the refs,  \listrefs
% (Extracted and adapted from harvmac.tex by P Ginsparg)

\catcode`\@=11 % This allows us to modify PLAIN macros.

\def\nolabels{\def\wrlabeL##1{}\def\eqlabeL##1{}\def\reflabeL##1{}}
\def\writelabels{\def\wrlabeL##1{\leavevmode\vadjust{\rlap{\smash%
{\line{{\escapechar=` \hfill\rlap{\sevenrm\hskip.03in\string##1}}}}}}}%
\def\eqlabeL##1{{\escapechar-1\rlap{\sevenrm\hskip.05in\string##1}}}%
\def\reflabeL##1{\noexpand\llap{\noexpand\sevenrm\string\string\string##1}}}
\nolabels
\global\newcount\refno \global\refno=1
\newwrite\rfile
\def\defref{$^{{\hbox{\rfont [\the\refno]}}}$\nref}
\def\nref#1{\xdef#1{\the\refno}\writedef{#1\leftbracket#1}%
\ifnum\refno=1\immediate\openout\rfile=refs.tmp\fi
\global\advance\refno by1\chardef\wfile=\rfile\immediate
\write\rfile{\noexpand\item{#1\ }\reflabeL{#1\hskip.31in}\pctsign}\findarg}
%	horrible hack to sidestep tex \write limitation
\def\findarg#1#{\begingroup\obeylines\newlinechar=`\^^M\pass@rg}
{\obeylines\gdef\pass@rg#1{\writ@line\relax #1^^M\hbox{}^^M}%
\gdef\writ@line#1^^M{\expandafter\toks0\expandafter{\striprel@x #1}%
\edef\next{\the\toks0}\ifx\next\em@rk\let\next=\endgroup\else\ifx\next\empty%
\else\immediate\write\wfile{\the\toks0}\fi\let\next=\writ@line\fi\next\relax}}
\def\striprel@x#1{} \def\em@rk{\hbox{}}
\def\lref{\begingroup\obeylines\lr@f}
\def\lr@f#1#2{\gdef#1{\defref#1{#2}}\endgroup\unskip}
\newwrite\lfile
{\escapechar-1\xdef\pctsign{\string\%}\xdef\leftbracket{\string\{}
\xdef\rightbracket{\string\}}}

\def\writestop{\def\writestoppt{\immediate\write\lfile{\string\p
ageno%
\the\pageno\string\startrefs\leftbracket\the\refno\rightbracket%
\string\def\string\secsym\leftbracket\secsym\rightbracket%
\string\secno\the\secno\string\meqno\the\meqno}\immediate\closeout\lfile}}
\def\writestoppt{}\def\writedef#1{}
\catcode`\@=12 % at signs are no longer letters

{\nopagenumbers
$~$\vskip 3truecm
\line{REVISED VERSION\hfill M/C--TH 93/11}
\line{\hfill DAMTP 93--23}
\vskip 20truemm
\centerline{{\fourteenrm PROTON STRUCTURE FUNCTION AT SMALL }{\fourteenit
Q}{$^{\hbox {2}}$}}
\bigskip
\bigskip
\centerline{A Donnachie$^{\dag}$}\footnote{}{$^{\dag}$ad@v2.ph.man.ac.uk}
\centerline{Department of Physics, University of Manchester}
\medskip
\medskip
\centerline{P V Landshof{}f$^*$}\footnote{}{$^*$pvl@amtp.cam.ac.uk}
\centerline{DAMTP, University of Cambridge}
\vskip 1truein
{\bf Abstract}

A fit is made to the data for the proton structure function up to $Q^2=10$
GeV$^2$, including the
real $\gamma p$ total cross-section. It is economical and simple, and
its form is motivated by physical principles. It is extrapolated down to
very small values of $x$. Data for the ratio $\nu W_2^n/\nu W_2^p$ are
also fitted.
\vskip 1truein
{\sl A FORTRAN program for the fit to $\nu W_2^p$ is available
by email on request}
\vfill
May 1993
\bigskip\eject}
\openup 4pt

The structure function $\nu W_2$ for inelastic electron or muon scattering
is constrained by gauge invariance to vanish linearly with $Q^2$ at
$Q^2=0$. Indeed, the total cross-section for real-photon scattering is
$$
\sigma (\gamma p)={{4\pi ^2\alpha _{\hbox{{\fiverm EM}}}}\over {Q^2}}\nu W_2
\eqno(1)
$$
evaluated at $Q^2=0$. For this reason, if for no other, the variation of
$\nu W_2$ with $Q^2$ at small $Q^2$ cannot be described by perturbative QCD:
it is unsafe to use any perturbative evolution equation until $Q^2$ is at
least so large that $\nu W_2$ has fully recovered from the need to vanish at
$Q^2=0$.

A few years ago\defref{\one}{
A Donnachie and P V Landshof{}f, Nuclear Physics B244 (1984)  322
} we parametrised the then-available data for $\sigma (\gamma p)$ and
for $\nu W_2$ at small $Q^2$ together in a very simple form.
A rather similar analysis has been repeated more recently\defref{\fourteen}{
H Abramowicz, E M  Levin, A  Levy, and U  Maor,
Physics Letters B269 (1991) 465;
G A Schuler and T Sj\"ostrand, preprint CERN-TH-6796/93
}.
Our parametrisation
compares quite well with more recent data, but we feel that the time has
now come to improve on it. This is for two reasons. The first is concerned
with the search for the Lipatov pomeron in the measurements of $\nu W_2$
at small $x$. There is some uncertainty about the effects of kinematic
constraints\defref{\two}{
J C Collins and P V Landshof{}f, Physics Letters B276 (1992) 196
}\defref\twoa{
J Forshaw, P Harriman and P Sutton, in preparation
}. and of shadowing\defref{\three}{
L Gribov, E M Levin and M G Ryskin, Physics Reports 100 (1983) 1
} on the solution to the Lipatov equation, and so it may not be immediately
obvious from future data whether the Lipatov pomeron is actually present.
To help decide this, it will be necessary to know as accurately as possible
what is expected from the soft pomeron, which is the pomeron
that is seen in all the data that have been collected so far\defref{\ffour}{
P V Landshof{}f, Proc XXVII Rencontre de Moriond,
(Editions Fronti\'eres, 1992, ed Tran Thanh Van)
}. Our new parametrisation allows what we believe to be a reliable
extrapolation to very small $x$ of the effect of soft pomeron exchange.
The second reason good fits to small-$Q^2$ data are needed is for
calculation of QED radiative corrections. These can be huge at HERA and so
they must be calculated as accurately as possible, using more than one
parametrisation of existing data as a check.

Our aim is to present a parametrisation of the data that is as simple
as possible, with rather few parameters, and whose component parts are
motivated by physical considerations.

We have shown recently\defref{\four}{
A Donnachie and P V Landshof{}f, Physics Letters B296 (1992) 227
} that all total cross-sections may be parametrised as a simple sum of
two Regge powers
$$
\sigma^{\hbox{{\sevenrm TOT}}}=Xs^{0.0808}+Ys^{-0.4525}
\eqno(2)
$$
We fixed the two powers in this expression from $pp$ and $\bar pp$ data, and
found that the resulting expression works well also for $\pi p$ and
$K p$. In the case of $\gamma p$, the fit (2) is very similar to one we
made a few years ago\ref{\one}, and it is in agreement with the published
data points from HERA\defref{\hera}{
ZEUS collaboration, Physics Letters B293 (1992) 456; H1 collaboration,
Physics Letters B299 (1993) 385
}. We show this in figure 1a. (Notice that although we have specified
the two powers to high accuracy, the data do not determine them to such
accuracy; the $pp$ and $\bar pp$ data from which we derived them would
allow slightly different values, with corresponding changes to the
coefficients $X$ and $Y$ -- the errors in the parameters are strongly
correlated.)

According to the parton model, the same Regge powers appear\defref{\five}{
P V Landshof{}f and J C Polkinghorne, Physics Reports 5C (1972) 1
} as powers of
$1/x$ in the small-$x$ behaviour of $\nu W_2$. The parton model is
drawn in figure 2, where the amplitude $T$ is the amplitude for the
emission of a parton of momentum $k$ by the proton.  From simple
kinematics\ref{\five}, the energy variable for this amplitude
is
$$
(p-k)^2=-x^{-1}(k^2+{\bf k}_T^2)-k^2+(1-x)m_p^2
\eqno(3)
$$
and so is large when $x$ is small.
Because $T$ is a hadronic amplitude, its high-energy behaviour should
involve the same Regge powers of $(p-k)^2$ as appear as powers of $s$
in (2), and these reflect themselves as corresponding powers of $1/x$
in $\nu W_2$. Of course the parton model is only the first term in a
perturbative-QCD expansion of $\nu W_2$;  the influence of the perturbative
corrections on the Regge powers is not understood and is the reason for
the great interest in small-$x$ physics at HERA. One possibility is that
the Regge powers in (2) survive the perturbative corrections and appear
in the small-$x$ behaviour, at least for small, and even perhaps moderately
large, values of $Q^2$. We explore this possibility in this paper.

We note that Monte Carlo models are commonly in error by taking
both $k^2$ and ${\bf k}_T$ to be zero.
While this may be
a good approximation for many purposes, it is not good when $x$ is small,
because, according to (3),
it fails to make $(p-k)^2$ large and so does not correctly take
account of the nonperturbative Regge behaviour.

Our first fit, then, is to the small-$x$ data from
NMC\defref{\six}{
NMC collaboration, Physics Letters B295 (1992) 159
}. We use two simple powers of $x$, each multiplied by a simple
function of $Q^2$ that vanishes linearly with $Q^2$ as $Q^2\to 0$
and goes to 1 for large $Q^2$:
$$
\nu W_2\sim Ax^{-0.0808}\left ({{Q^2}\over{Q^2+a}}\right )^{1.0808}
+Bx^{0.4525}\left ({{Q^2}\over{Q^2+b}}\right )^{0.5475}
\eqno(4a)
$$
with the constraints that
$$
Aa^{-1.0808}=0.604~~~~~~~~~~~~~~~~~~~Bb^{-0.5475}=1.15
\eqno(4b)
$$
so as to retrieve the fit of figure 1a to the real-photon data when
$Q^2\to 0$.
This two-parameter fit is compared with the NMC data in figure 3,
for the choices $A=0.324$ and $B=0.098$. Our fit has used only the data
up to $Q^2=10$.

Strictly speaking, because $x=Q^2/2\nu$,
the $Q^2\to 0$ limit of (4) is not a sum of powers
of $s$ as in (2), but rather a sum of powers of $2\nu =s-m^2_N$.
At the lowest $s$-value of the data in figure 1a the difference is negligible,
though if we continue to smaller $s$ it becomes important.
We show the curve obtained for $\sigma (\gamma p)$ from the
$Q^2\to 0$ limit of (4) in figure 1b. Even though we have not used the data
below $\surd s=6$ to determine the fit, it works satisfactorily down into
the region where the resonances begin to be important.

If we extrapolate (4) to very small $x$, we obtain a predicted value just less
than 0.8 for $\nu W_2$ at $x=10^{-5}, Q^2=10$.
This prediction remains rather stable
under the refinements to our fit that we report below.
We make refinements to the fit partly in order to extend it to larger
values of $x$, as is necessary if it is to be useful for making radiative
corrections. But also we want to incorporate the contribution from heavy
flavours into the analysis, since part of the rapid rise with $Q^2$
seen in the data in figure 3 is to be attributed to their very rapid
switching-on.

The charmed-quark contribution to $\nu W_2$ in muon scattering has been
measured\defref{\seven}{
BFP collaboration, Physical Review Letters 45 (1980) 1465;
EMC collaboration, Nuclear Physics B213 (1983) 31
} and found to vary extremely rapidly with $Q^2$.
It seems
that this is a threshold effect, though we must emphasise that there is no
good theoretical understanding of how one should take account of threshold
effects\defref{\nine}{
P V Landshof{}f and D M Scott, Nuclear Physics B131 (1977) 172;
V Barone, M  Genovese, N N Nikolaev, E  Predazzi and
B Z  Zakharov, Physics Letters B268 (1991) 279 and  B304 (1993) 176
}. Such an understanding would require more knowledge than we have of the
effects of confinement. Nevertheless, a successful phenomenology of the data
is available\ref{\seven}\defref{\eight}{
A Donnachie and P V Landshof{}f, Physics Letters B207  (1988)  319
}: the threshold effect is well described by supposing that the contribution
$F_2^{c\bar c}$ to $\nu W_2$ from charmed quarks is a function not of the
Bjorken variable $x$ but of
$$
\xi _c=x\left (1+{{\mu _c^2}\over{Q^2}}\right )
\eqno(5a)
$$
Then
$$
(1-\xi _c)={{W^2-W_0^2}\over{2\nu}}
\eqno(5b)
$$
with $W_0^2=\mu _c^2+m_p^2$, and
the best description of the data corresponds to choosing the parameter $\mu _c$
such that $W_0=m_D+m_{\Lambda _c}$, the threshold value of $W$. (Strictly
speaking, the threshold is different according to whether it is a quark
or an antiquark that has absorbed the virtual photon, but the data are
not sufficiently accurate for this to matter, and we simply take the lowest
physical threshold.)  Hence, by using
the variable $\xi _c$ instead of $x$, we ensure that $F_2^{c\bar c}$
goes to zero at threshold for each $Q^2$. According to the spectator-counting
rule\defref{\ten}{
J Gunion, Physical Review D10 (1974) 242;
R Blankenbecler and S J Brodsky, Physical Review D10 (1974) 2973
}
$F_2^{c\bar c}$ should behave as $(1-\xi _c)^7$ as $\xi_c\to 1$, while
Regge theory requires it to have power behaviour close to  $\xi _c^{-0.08}$
at small $\xi _c$. In addition, it must vanish linearly with $Q^2$ when
$Q^2\to 0$, which we achieve as in (4a). This led us in reference
{\eight} to the fit
$$
xc(x,Q^2)={\textstyle{{9\over 8}}}F_2^{c\bar c}=C_c {{Q^2}\over{Q^2+6.25}}\xi
_c^{-\epsilon}
(1-\xi _c)^7
\eqno(5c)
$$
where we took $\epsilon$ to be 0.086 and $C_c=0.045$.
For the reasons we have explained, we now adopt the value
0.0808 for $\epsilon$, but the difference is negligible.
Since we made this fit, the data have been
scaled downwards, because of a revised experimental value of the branching
ratio of the charmed quark decaying to a muon; we therefore shall
use the value $C_c=.032$. The fit, with the renormalised data, is shown
in figure 4.

Some discussion of the use of $\xi_c$ is called for. As we have indicated,
it is suggested by the data.  More naive theoretical
considerations might lead to the ``slow rescaling'' choice $\mu _c
=m_c$, the charmed
quark mass, instead of the 2 GeV that is needed. To derive such slow
rescaling  one would need
two assumptions: (i) that the mass scale associated with
the fragmentation of the charmed quark after it has absorbed the virtual
photon is $m_c$ rather than that of the lightest hadron to which it can
fragment; and (ii) one can ignore the fact that the momentum $k$ of the quark
before it absorbs the photon is not on shell. The fact that $\mu _c$ needs
to be so large indicates that neither of
these assumptions is tenable\ref{\five}:
not only is the minimum mass of the hadron to which the quark fragments
important, so also is the minimum mass of the residual fragments of the
proton left behind when the quark has been pulled out of it. The only way
that such a mass can be supported is for the quark momentum $k$ to go
off shell and become negative: see (3).

Given that the phenomenology of $F_2^{c\bar c}$ points to the use of
the variable $\xi _c$ defined in (5a), it is natural to assume that the
strange antiquark distribution $x{\bar s}(x,Q^2)$ should
be handled similarly. The corresponding
variable $\xi _{\bar s}$ should be defined
similarly, but with a scale $\mu _{\bar s}$
such that the threshold $W_0=m_K+m_{\Lambda}$. From (5b), this requires
$\mu _{\bar s}^2=1.7$ GeV$^2$. We shall work with
$$
x{\bar s}(x,Q^2)=C{{Q^2}\over{Q^2+a_s}}\xi _s^{-0.0808}(1-\xi _s)^7
\eqno(6)
$$
The threshold for the strange quark distribution $xs(x,Q^2)$ is somewhat
higher, but we shall not take account of this refinement,
because we also have to ignore the possibility that the
$s$ and $\bar s$ quarks recombine\defref{\recombine}{
A Donnachie and P V Landshoff, Nuclear Physics B112 (1976) 233
} and do not appear in the final state,
with then a much smaller value for $W_0$ and hence for $\mu _s$.
We can only guess what to take for the parameter $a_s$; we choose the value
1 GeV$^2$, which is in between the corresponding value for light quarks
(see (4)) and charmed quarks. Fortunately, its precise value will
not be too important for us.
Our first guess for the value of the
constant $C$ is that it is the same as for light quarks; (4) then gives
$C\approx 0.22$. We shall confirm from our later fit that this is the
appropriate value in the case of the light quarks, but of course it is
far from obvious that we should use it also for strange quarks. There is
some confusion about the magnitude of the strange-quark content of the proton.
A direct measurement from two-muon events in neutrino
scattering\defref{\thirteen}{
CCFR collaboration, Physical Review Letters 70 (1993) 134
} finds that the strange quark distribution is half as large as the light
antiquark distributions, but the less direct method\defref{\eleven}{
CTEQ collaboration, preprint FERMILAB-PUB-92-371
} of measuring
$$
xs(x,Q^2)\approx \textstyle{5\over 6}F_2^{\nu N}-3F_2^{\mu D}
\eqno(7)
$$
gives a result that is significantly larger\defref{\twelve}{
A.D.Martin, W.J.Stirling and R.G.Roberts,
Durham preprint DTP/93/22, RAL preprint RAL-93-027
(Proc Durham Workshop on HERA Physics, J Phys G, to be published)
}, even allowing for that fact
that it is risky to measure a small quantity as the difference between
two large ones, as in  (7).
As a contribution to this debate, we point out that the threshold effects,
being sensitive to the minimum mass that can be produced in the final
state, will vary according to how one measures $xs(x,Q^2)$.
In $F^{\nu N}_2$ there are two contributions. In the first, which is
Cabbibo suppressed, the strange quark absorbs the weak current and
becomes a light quark, so that the threshold is $W_0 = m_{\pi} +
m_{\Lambda}$ corresponding to $\mu_s$ = 0.8 GeV. In the second, the
strange quark absorbs the weak current and becomes a charm quark, so
that the threshold is $W_0 = m_D + m_{\Lambda}$ corresponding to $\mu_s$
= 2.9 GeV. The latter process is also the source of the dimuon events in
neutrino interactions, with the same value of $\mu_s$ (a much larger value
than was assumed by the experimentalists\ref{\ten} in the analysis of their
data!).
In $F^{\mu D}_2$ the strange quark survives and $W_0 = m_K + m_{\Lambda}$,
with then $\mu_s$ = 1.3 GeV. In the latter case there is, however,
the possibility that the $s$ and $\bar s$ quarks recombine with consequently
a smaller value for $\mu_s$.
As an additional complication, the requirement that $\nu W_2$
vanishes as $Q^2\to 0$ applies only to photon-exchange processes and not
to processes involving the non-conserved weak current. To some extent one
can take this into account\ref{\eight} through PCAC, but only for light-quark
distributions.

Having decided, albeit somewhat tentatively, how to handle the contributions
to $\nu W_2$ from the heavy quarks (fortunately they are small),
we now turn to the $u$ and $d$ quarks
and their antiquarks. With the threshold $W_0=m_{\pi}+m_p$, $\mu$=0.53 GeV,
and we use this to define a variable $\xi$ similar to (5a).
When $Q^2\to 0$, $\xi \sim \mu ^2/2\nu$, so that in order to have $\nu W_2$
vanishing linearly with $Q^2$ we replace the small-$x$ behaviour (4) with
$$
\nu W_2\sim A\xi ^{-0.0808}\phi (Q^2)
+B\xi ^{0.4525}\psi (Q^2)
\eqno(8a)
$$
where
$$
\phi (Q^2)={{Q^2}\over{Q^2+a}}~~~~~~~~~~~~
\psi (Q^2)={{Q^2}\over{Q^2+b}}
\eqno(8b)
$$
and
$$
Aa^{-1}(\mu ^2)^{-0.0808}=0.604~~~~~~~~~~~~~~~~~~~Bb^{-1}(\mu ^2)^{0.4525}=1.15
\eqno(8c)
$$
This is the behaviour we want when $x$ or $\xi$ is small; when $\xi$ is
close to 1 we impose instead the spectator-counting rules\ref{\ten} to
determine
the powers of $(1-\xi )$. However, we find that if we simply multiply the terms
in (4a) by such powers, they spoil the good fit of figure 3 to the small-$x$
data. This is because, while at the smallest $x$-value 0.008 of the data
in the figure, $(1-x)^7$ is close to 1, at the largest $x$-value of 0.07
it is already as small as 0.6, and the effect of using instead $(1-\xi )^7$
is even more marked.
In order to overcome this problem, we shall use the simple sum of powers
for values of $\xi$ less than some fixed value $\xi _0$, which we leave
as a free parameter. For $\xi > \xi _0$ we match to each term in (8a)
expressions of the form
$$
\hbox{const }\xi ^{\lambda} (1-\xi )^m
$$
where the power $\lambda$ is fixed by requiring such a form to fit smoothly
on to the simple power of $\xi$ at $\xi =\xi _0$, that is the two forms
and their first derivatives are equal there. We shall use the same value
of $\xi _0$ throughout, and find that the best fit requires it to be quite
small, about 0.07.

Consider first the valence distribution $xu_V(x,Q^2)$.
At small  $\xi$ it should have the power behaviour 0.4525, corresponding
to $\omega /\rho$ exchange. As $\xi\to 1$ its behaviour should be
$(1-\xi )^3$, according to the spectator-counting rule\ref{\ten}. Hence
we take
$$
xu_V(x,Q^2)=U(\xi )\psi (Q^2)
$$
$$
U(\xi )=\cases{B_u\xi ^{0.4525} &$\xi <\xi _0$\cr
                  \beta _u \xi ^{\lambda _u}(1-\xi )^3 & $\xi >\xi _0$\cr}
\eqno(9a)
$$
We fix $\beta _u$ and $\lambda _u$ in terms of $B_u$ by requiring these
two forms to join smoothly at $\xi =\xi _0$, and then determine
$B_u$ by imposing the number sum rule
$$
\int _0^1 {{d\xi}\over{\xi}}U(\xi )=2
\eqno(10a)
$$
The valence distribution $xd_V(x,Q^2)$ should have the same power
behaviour 0.4525. The spectator-counting rule would make it also behave as
$(1-\xi )^3$ as $\xi\to 1$; however, it is well known that this conflicts
with the measured ratio\defref{\fifteen}{
NMC collaboration, Nuclear Physics NP B371 (1992) 3
} $\nu W_2^n/\nu W_2^p$, which indicates that in
$xd_V$ the coefficient of $(1-\xi )^3$ is very small.
So we take
$$
xd_V(x,Q^2)=D(\xi )\psi (Q^2)
$$
$$
D(\xi )=\cases{B_d\xi ^{0.4525} &$\xi <\xi _0$\cr
                  \beta _d \xi ^{\lambda _d}(1-\xi )^4 & $\xi >\xi _0$\cr}
\eqno(9b)
$$
with $\beta _d$ and $\lambda _d$ again being fixed in terms of $B_d$ by
joining the two forms smoothly and then $B_d$ determined from
$$
\int _0^1 {{d\xi}\over{\xi}}D(\xi )=1
\eqno(10b)
$$

For the nonvalence distributions, for each light quark and antiquark we
include a term that behaves as $\xi ^{-0.0808}$ for small $\xi$:
$$
\cases{C \xi ^{-0.0808}\phi (Q^2)&$\xi <\xi _0$\cr
           \gamma \xi ^{\lambda _s}(1-\xi )^7 \phi (Q^2)& $\xi >\xi _0$\cr}
\eqno(11a)
$$
The constants $\gamma$ and $\lambda _s$ are fixed in terms of $C$ by joining
these two forms smoothly at  $\xi = \xi _0$, with the same $\xi _0$ as
for the valence terms. There is no number sum rule for the nonvalence
distributions, so $C$ is a free parameter. In $\nu W_2$, (11a) is
multiplied by $10/9$, that is the constant $A$ in (8a) is $10C/9$.

When the two-component parton model was first formulated\defref{\sixteen}{
P V Landshof{}f and J C Polkinghorne, Nuclear Physics B28 (1971) 225
}, the component that later came to be called nonvalence\defref{\seventeen}{
J Kuti and V F Weisskopf, Physical Review D4 (1971) 3418
} had no contribution from $\rho,\omega,f,a_2$ exchange. This is because, at
that
time, the idea of exchange degeneracy was taken very seriously, not just for
the Regge trajectories but also for their couplings.  Since then, it has become
clear that the degeneracy of the trajectories is satisfied very well but
there is no basis for supposing that it extends also to the
couplings\ref{\four}. So we include in each
nonvalence distribution also a term
behaving as $\xi ^{0.4525}$ at small $\xi$.
We write the total such contribution to
$\nu W_2^p$ as
$$
\cases{(B-B_u-B_d)\xi ^{0.4525}\psi (Q^2)&$\xi <\xi _0$\cr
           \beta \xi ^{\lambda}(1-\xi )^9 \psi (Q^2)& $\xi >\xi _0$\cr}
\eqno(11b)
$$
where again there is only one free parameter, $B$, after we have joined the
two forms smoothly. Notice that we have not specified how this term divides
among the quark flavours: while by definition (because the term is nonvalence)
$u$ and $\bar u$ receive equal contributions, as do $d$ and $\bar d$,
it is likely\defref{\eighteen}{
A D Martin, R G Roberts and W J Stirling, preprint RAL-92-078, Physics
Letters B (in press)
} that $\bar u \not= \bar d$. Notice also that we have, somewhat arbitrarily,
chosen the power $(1-\xi )^9$; the spectator-counting rule gives no guidance
here, but we need a power greater than 7 to prevent the nonvalence distribution
from becoming negative for large values of $\xi$, since while $B_u$ is
close to $\textstyle {3\over 2}$ and $B_d$ is about half that,
the best fit for $B$ is at quite a small
value, as we already saw in figure 3. Again somewhat arbitrarily, we have
used the same factor $\psi (Q^2)$ to make all the $\xi ^{0.4525}$ terms vanish
linearly with $Q^2$ at small $Q^2$.

The terms we have included so far have the the property that, at each $x$,
$\nu W_2$ increases as $Q^2$ increases, while the data show the opposite
when $x$ is not small. At large $Q^2$ this is a consequence of the
perturbative evolution, but we are constructing a fit at $Q^2$-values
where perturbation theory is not applicable. We therefore include an
additional ``higher-twist'' term
$$
ht(x,Q^2)=D{{x^2(1-\xi )^2}\over{1+Q^2/Q_0^2}}
\eqno(12)
$$
This resembles a term we introduced some years ago\defref{\nineteen}{
A Donnachie and P V Landshof{}f, Physics Letters 95B (1980) 437
}, when we indentified it as a contribution from the virtual photon
being absorbed by a diquark within the proton, though we do not necessarily
adhere to this interpretation now. By making it vanish
quadratically as $x\to 0$, we have ensured that it does not contribute
to the real-photon cross-section, while the power $(1-\xi )^2$ gives a better
fit than would $(1-\xi )$. This term has two free parameters, $D$ and $Q_0$,
in addition to the $\xi _0$, $C$ and $B$ we already have. Our best fit to
the data for $Q^2<10$ GeV$^2$ corresponds to
$$
C=0.220\;~~~~  B=0.279\;~~~~ \xi_0=0.071\;~~~~ D=15.88\;~~~~Q_0=550 \hbox{ MeV}
\eqno(13a)
$$
and is shown in figure 5. The $\chi ^2$ per data point is just less than
0.5. Again we have quoted the values of the parameters to high
accuracy because the errors, particularly in $D$ and $Q_0$,
are strongly correlated. For example,
$$
C=0.220\;~~~~ B=0.274\;~~~~ \xi_0=0.075\;~~~~ D=36.74\;~~~~Q_0=338 \hbox{ MeV}
$$
gives almost exactly the same $\chi ^2$. The values we obtain for $C$ and
$B$ are rather stable, particularly that for $C$.
The values of the subsidiary parameters corresponding to the choices (13a) are
$$
B_u=1.456~~~~ B_d=0.772~~~~ \lambda _u=0.683~~~~
\lambda _d=0.760~~~~ \lambda=1.144~~~~ \lambda _s=0.457
\eqno(14)
$$

We emphasise that our nonperturbative fit is supposed only to apply
for $Q^2<10$. For larger values of $Q^2$ it lies above the data at moderate
and large
$x$, allowing room for the perturbative evolution to take over at $Q^2=10$,
or at some smaller value. See figure 6. We remark also that our choice to
fit the data up to $Q^2=10$ is somewhat arbitrary. If we replace this
with $Q^2=5$ the values of the parameters hardly change:
$$
C=0.213 \;~~~~ B=0.312\;~~~~\xi_0=0.069\;~~~~ D=15.88\;~~~~  Q_0=554 \hbox{
MeV}
$$
and the $\chi ^2$ per data point is reduced to 0.3.

We have extended our fit to $\nu W_2^n$, though the data here are rather
more uncertain. Not only is there a lack of basic knowledge on exactly how
to make deuterium corrections\defref{\twenty}{
P V Landshof{}f and J C Polkinghorne, Physical Review D18 (1978) 153;
B Badelek and J Kwiecinski, Nuclear Physics B370 (1992) 278;
M Strikman, XXVI International Conference on High Energy Physics (1992)
}, but also the NMC data are in the course of being changed
slightly\defref{\*one}{
B Badelek, talk at Durham Workshop, March 1993
}. We use the NMC data, as published\ref{\fifteen}, for
$\nu W_2^n/\nu W_2^p$. Our fit to $\nu W_2^n$ uses the same values for
$\xi _0$ and $C$ as for $\nu W_2^p$, but allows for a different $B$,
corresponding to $\bar u \not=\bar d$, and different ``higher twist''.
We choose to work with the same mass scale $Q_0$ in the latter, but allow
its normalisation $D$ to be chosen by the fit to the data.
So we have 2 new free parameters.
For $\nu W_2^n$ we end up with
$$
C=0.220\;~~~~  B=0.169\;~~~~ \xi_0=0.071\;~~~~ D=4.94\;~~~~  Q_0=550 \hbox{
MeV}
\eqno(13b)
$$
which gives the fit shown in figure 7. Once again, the data do not determine
the ``higher twist'' term at all precisely: we could have used a very different
value of $D$ with $Q_0$ changed correspondingly. But it is clear that we
need different terms
for $\nu W_2^p$ and $\nu W_2^n$, not only for the ``higher twist'' but
also for the nonvalence $\xi ^{-0.0808}$ term:
for $\nu W_2^p$ its coefficient
is -0.454, while for $\nu W_2^n$
it is -0.337, so that indeed $\bar u \not=\bar d$.

We have succeeded in obtaining an excellent description of $\nu W_2^p$ and
$\nu W_2^n/\nu W_2^p$ in the range $0<Q^2<10$ with very few parameters.
The form of our fit is simple, and motivated by theoretical principles.
Our analysis should be regarded as complementary to the standard
ones\ref{\eleven}\ref{\eighteen}, which are mainly driven by considerations
of perturbative QCD.
We do not use any neutrino-scattering data to determine the free parameters.
As we have explained, until $Q^2$ is rather larger than the range with which
we are concerned, the neutrino data are not expected to be related  at all
precisely to the photon-exchange data, beacuse they approach the limit
$Q^2\to 0$ differently. On the other hand, we fit the photon-exchange data
down to small $Q^2$, even $Q^2=0$.
The urgent question still remains of how properly to combine the
perturbative and nonperturbative analyses,
such as has been attempted for example
by Badelek and Kwiecinski\defref{\*two}{
B Badelek and J Kwiecinski, Physics Letters B295 (1992) 263
}.

We have already remarked that the extrapolation of our fit down to very
small $x$ is finally not very different from that obtained from (4).
In figure 8 we show our curves plotted against $x$. If the HERA experiments
find results for $\nu W_2$ significantly larger at small $x$ than our
extrapolations, we claim that this will be a clear signal that they have
discovered new physics. Of course, the hope is that they will discover the
Lipatov pomeron. In relation to this, the question arises whether
the effect of the soft pomeron, which is included in our curves, should
be subtracted off from the data\ref{\two}, with the
Lipatov pomeron being identified
with anything that may remain, or whether instead the Lipatov pomeron
simply {\it replaces} the soft pomeron in the small-$x$ behaviour of $\nu W_2$.
This is all part of the general study of the interface between
perturbative and nonperturbative QCD, which will be so important at HERA.
\bigskip
{\sl One of us (PVL) is pleased to acknowledge vigorous discussions with
Jean-Ren\'e Cudell, Steve Ellis and Dieter Haidt.}

\vfill\eject
\medskip\immediate\closeout\rfile\writestoppt
\baselineskip=14pt{{\bf References}}\bigskip{\frenchspacing%
\parindent=20pt\escapechar=` \input refs.tmp\bigskip}\nonfrenchspacing
\def\h{\hfill\break}
\vfill\eject

{\bf Figure captions}
\medskip
\i{1} Data for the real-photon cross-section, with the fit from reference
\four.

\i{2} The parton model

\i{3} NMC data at small $x$, with simple-power fit.
Reading from top to bottom, the values of $x$ are\h
\centerline{0.008, 0.0125, 0.0175, 0.025, 0.035, 0.05, 0.07}\h
For clarity of presentation, the curves and data have been scaled by a
different
factor at each value of $x$; the scale factors are\h
\centerline{4, 3.2, 2.5, 2, 1.5, 1.2, 1}

\i{4} EMC data for $xc(x,Q^2)$ with fit from reference \eight. The data
have been renormalised to take account of the revised branching ratio
for a charm quark decaying to a muon.
Reading from top to bottom, the values of $x$ are
0.00422,\ 0.0075,\ 0.0133,\ 0.0237,\ 0.0422,\ 0.075,\ 0.133,\ 0.237~.
For clarity of presentation, the curves and data have been scaled by a
different
factor at each value of $x$; the scale factors are 128, 64, 32, 16, 8, 4, 2, 1
respectively.

\i{5} Data from NMC, SLAC and BCDMS with fit described in the text.
For clarity of presentation, the curves and data have been scaled by a
different
factor at each value of $x$.
Reading from top to bottom, the values of $x$, with the scale factors in
brackets, are

{\leftskip 2truecm
{\ii{(a)} 0.008 (5), 0.0125 (4), 0.0175 (3.2), 0.025 (2.5), 0.035 (2),
0.05 (1.5), 0.07 (1)
\ii{(b)} 0.09 (7), 0.1 (5), 0.11 (3.5), 0.14 (2.5), 0.18 (2), 0.225 (1.5),
0.275 (1)
\ii{(c)} 0.35 (32), 0.45 (16), 0.5 (8), 0.55 (4), 0.65 (2), 0.75 (1)}

}

\i{6} The curves of figure 6b extrapolated to larger $Q^2$.
Reading from top to bottom, the values of $x$ are 0.09, 0.18, 0.275,
0.45 and 0.65, with scale factors 8, 4, 2, 1, and 1.

\i{7} NMC data for $\nu W_2^n/\nu W_2^p$, with fit. The data and curve
correspond to values of $Q^2$ that vary with $x$, from $\langle Q^2\rangle$
=0.4 GeV$^2$ at the smallest $x$ to 10.8 at the largest.

\i{8} Fits to $\nu W_2^p$ extrapolated to very small $x$ for three values
of $Q^2$. The curves
are the total, together with its valence, nonvalence and ``higher-twist''
components

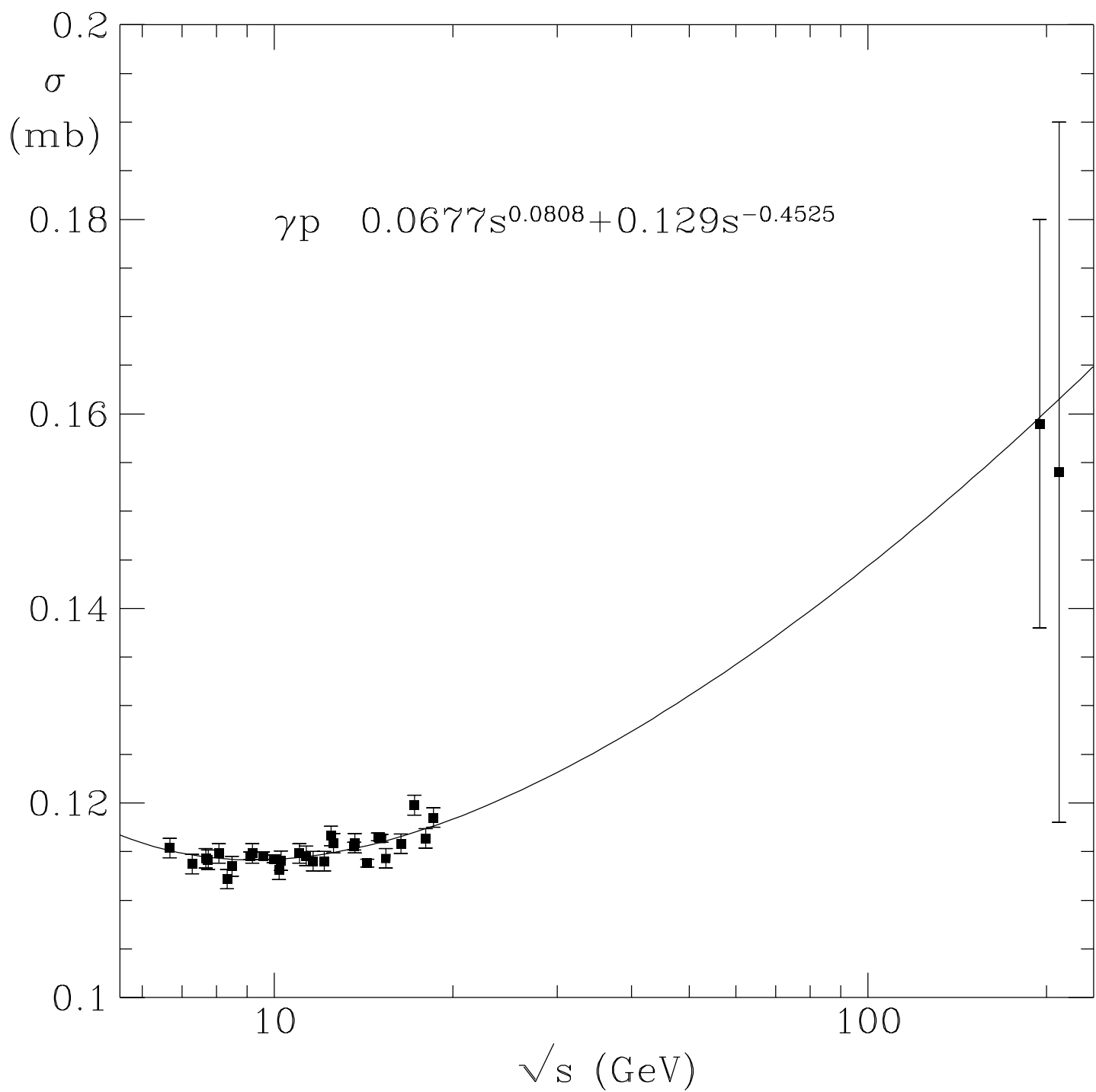
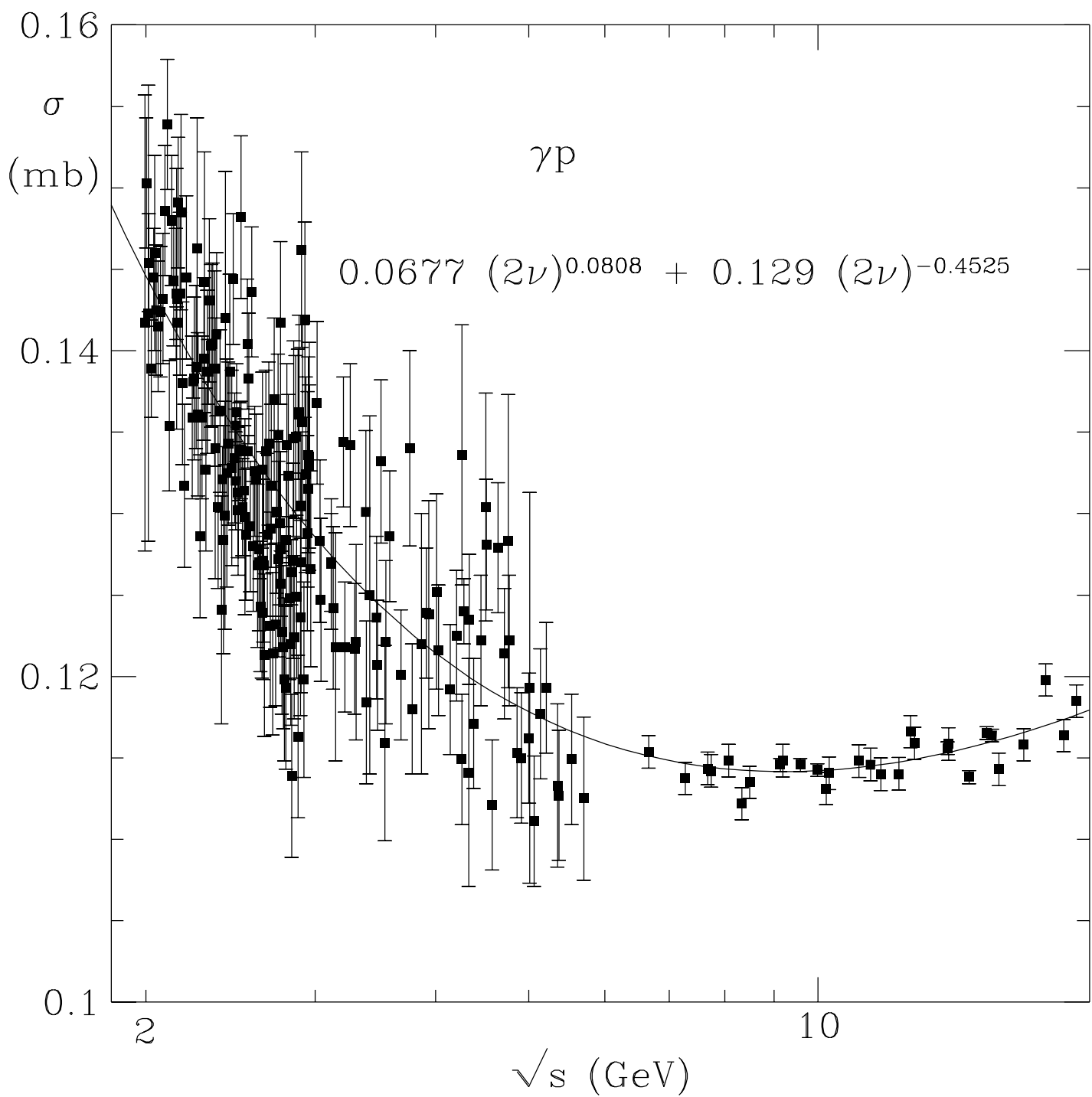
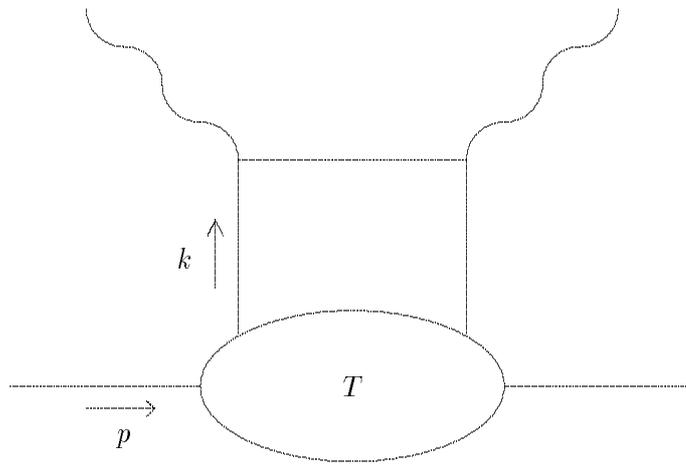
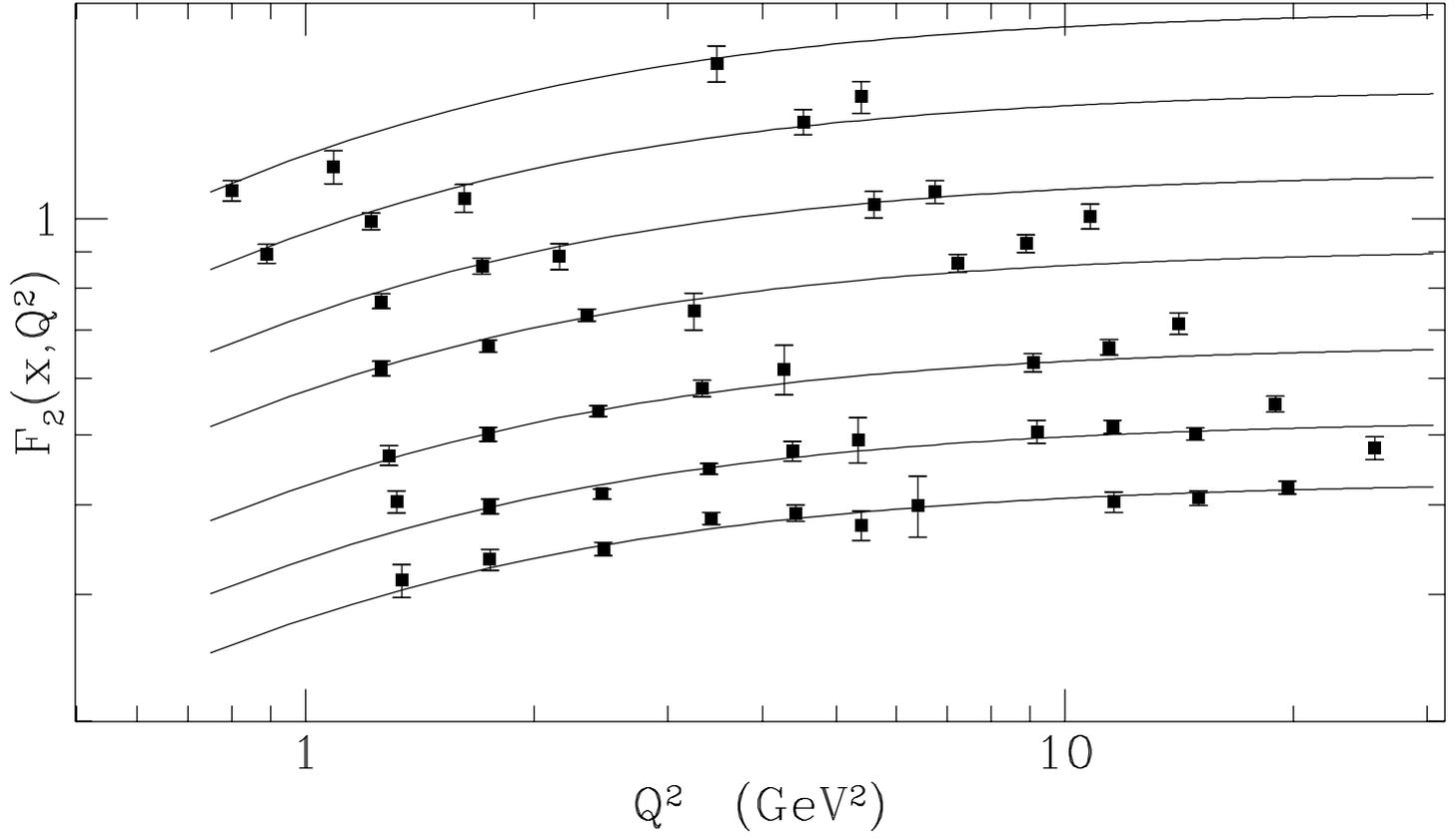
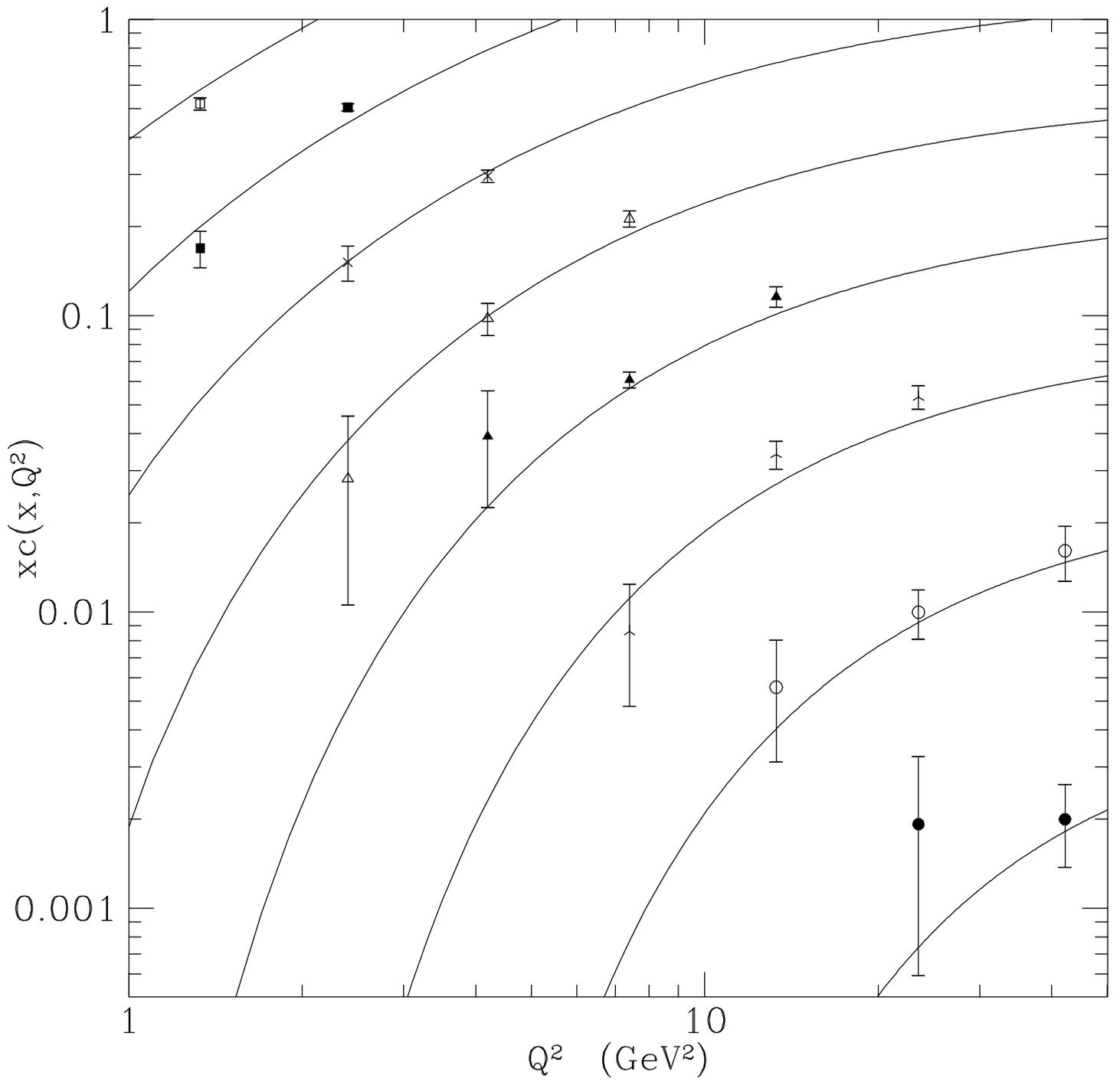
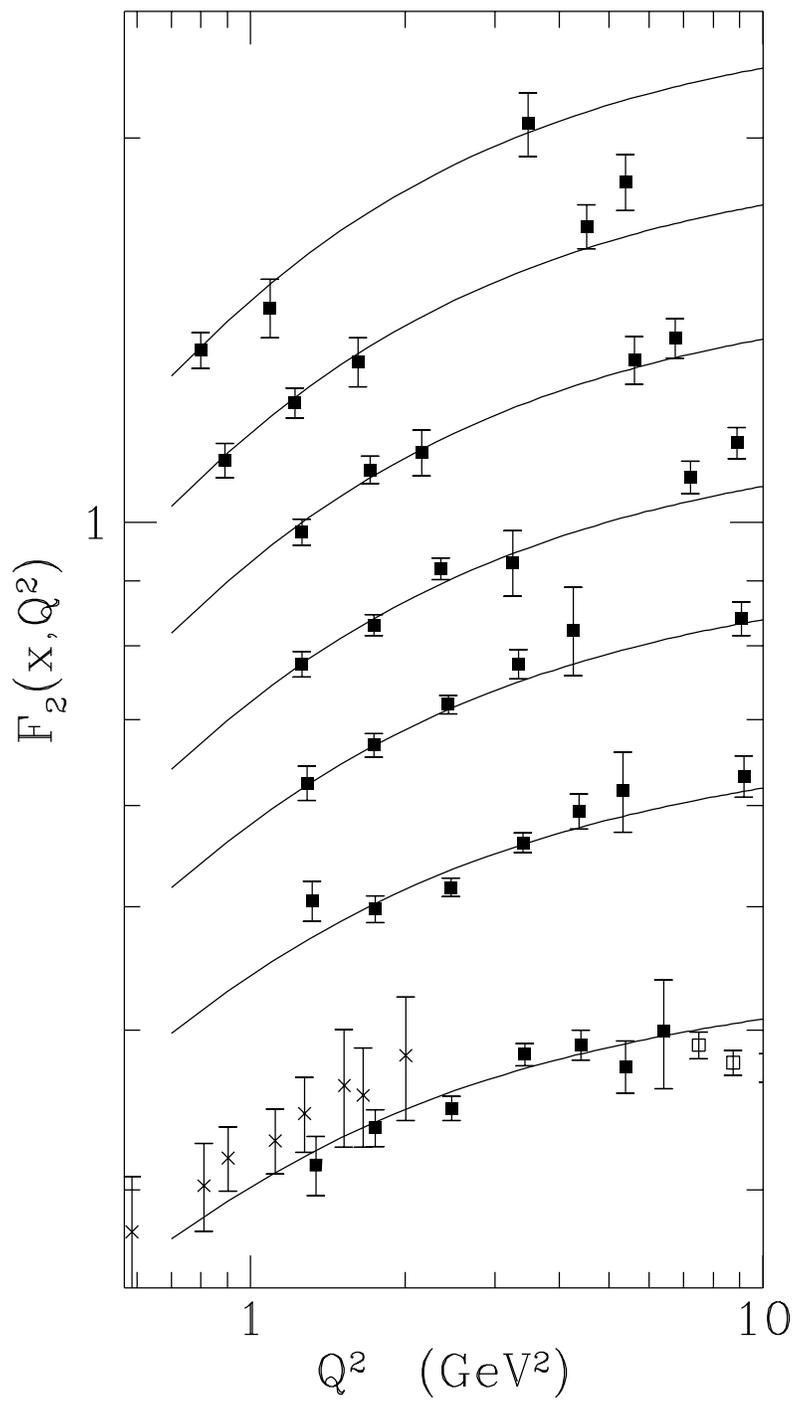
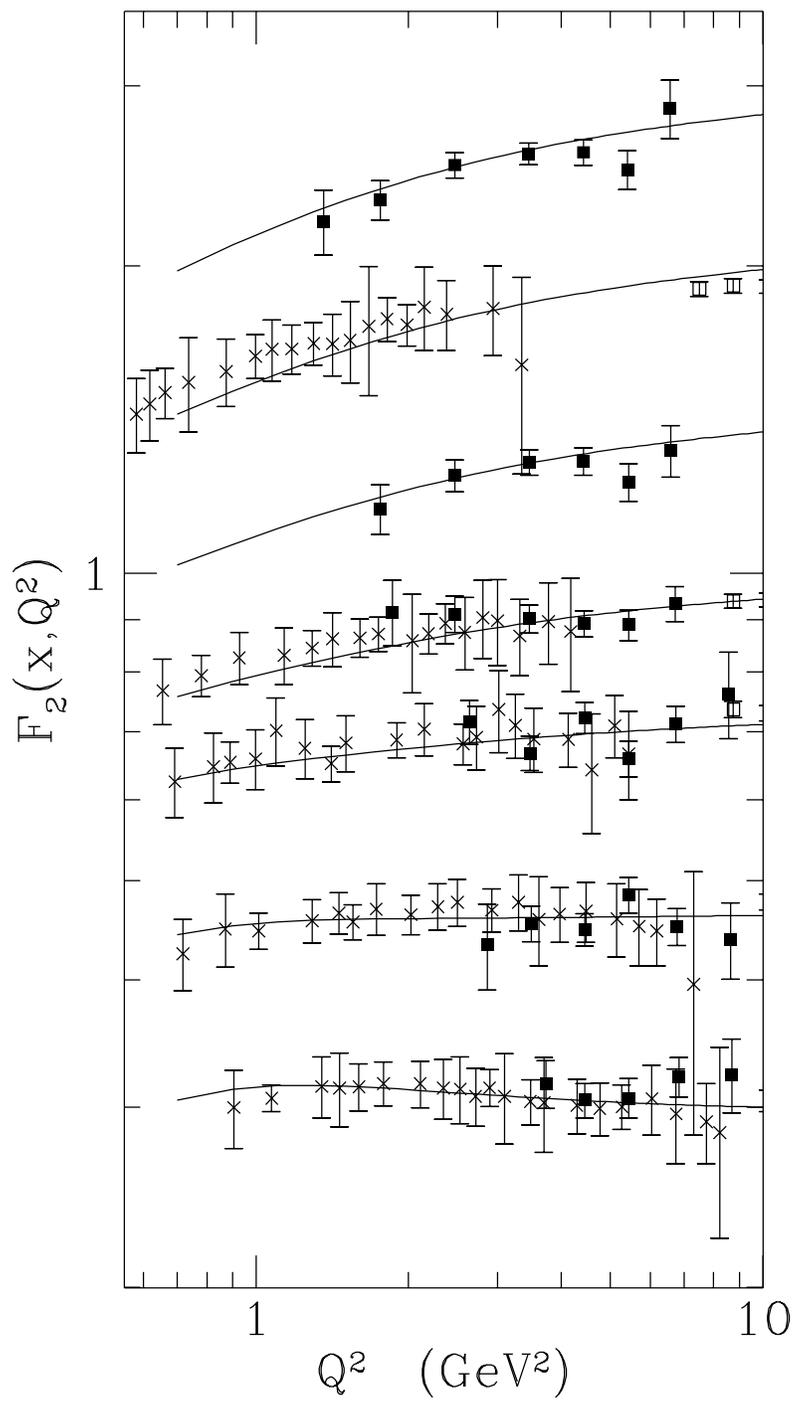
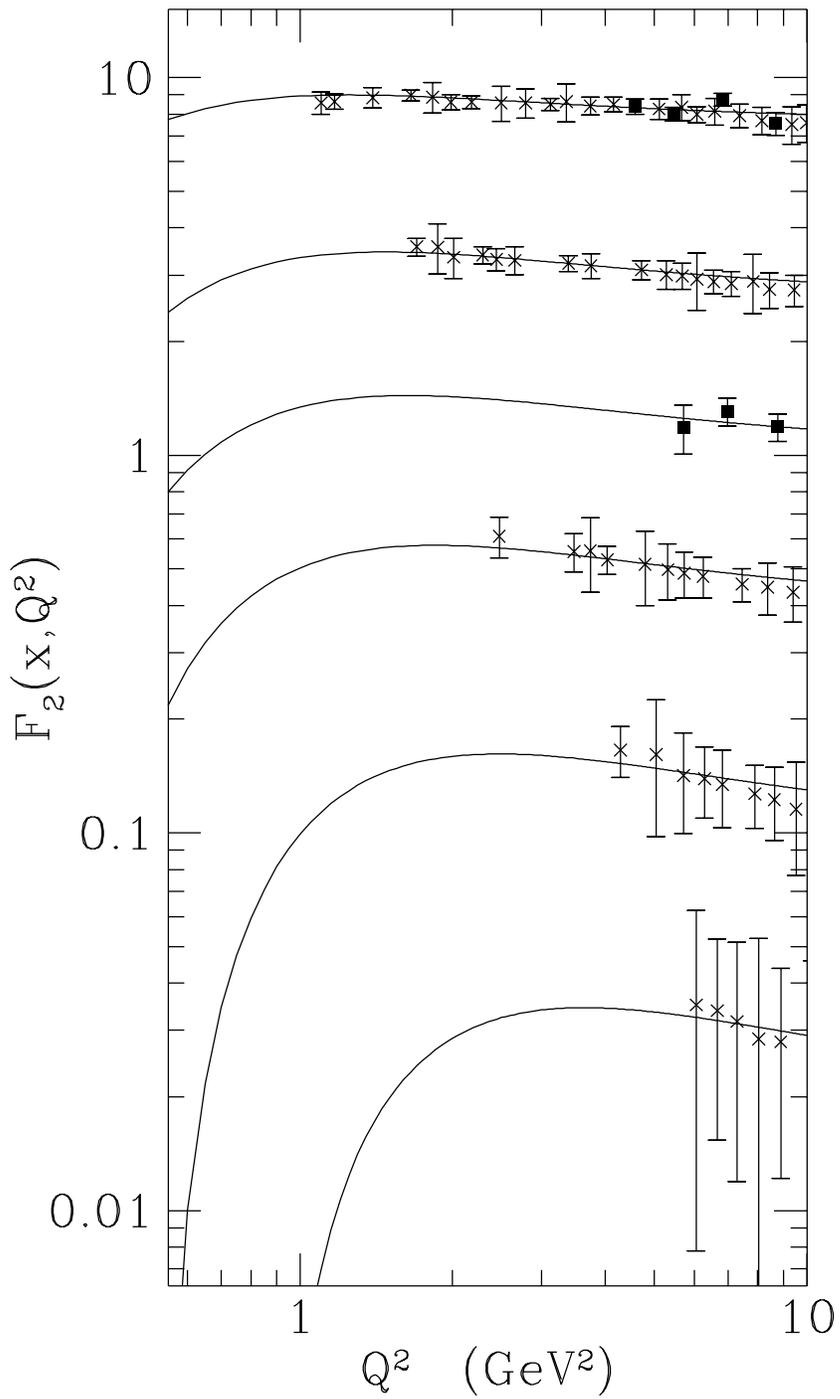
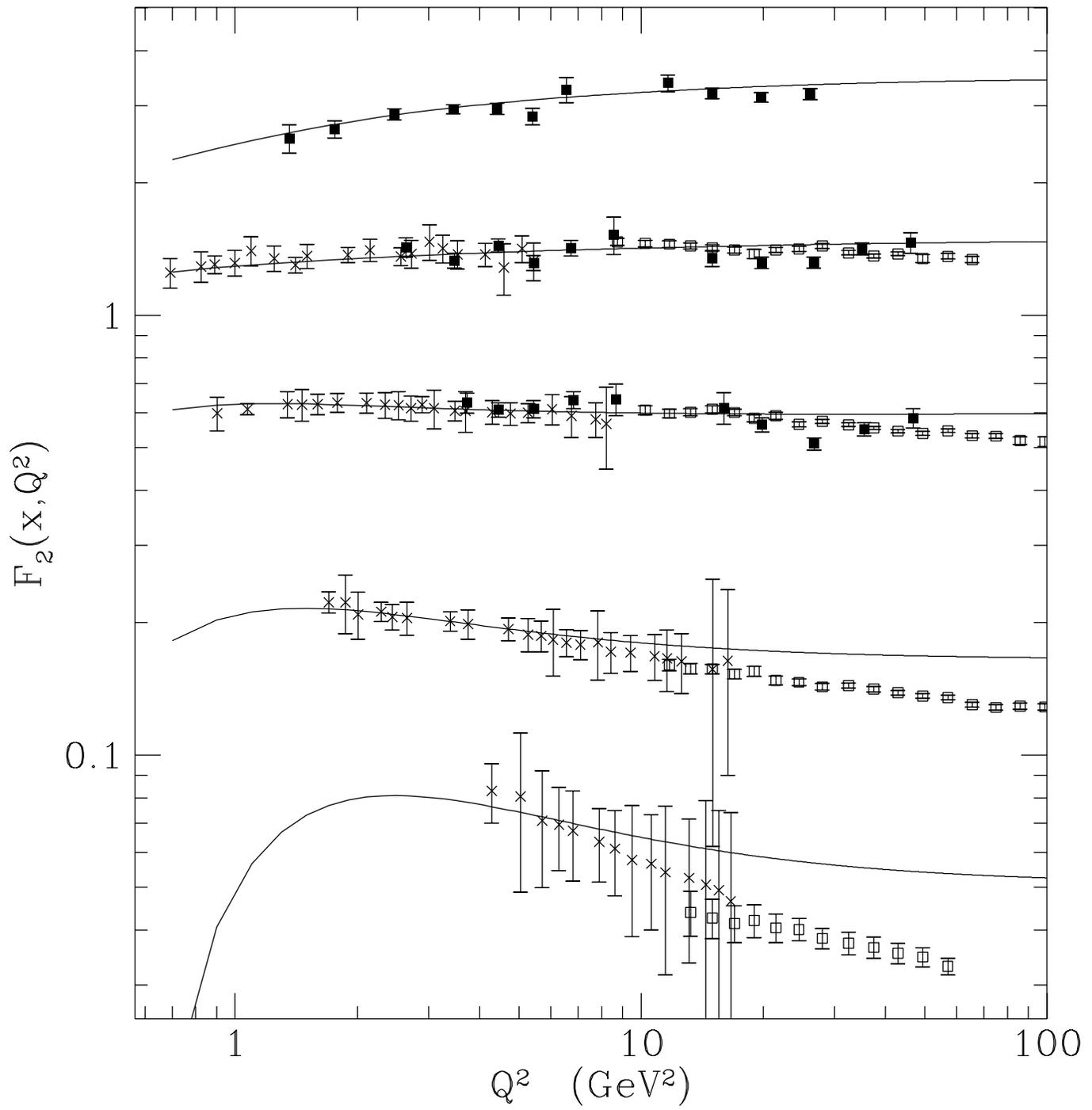
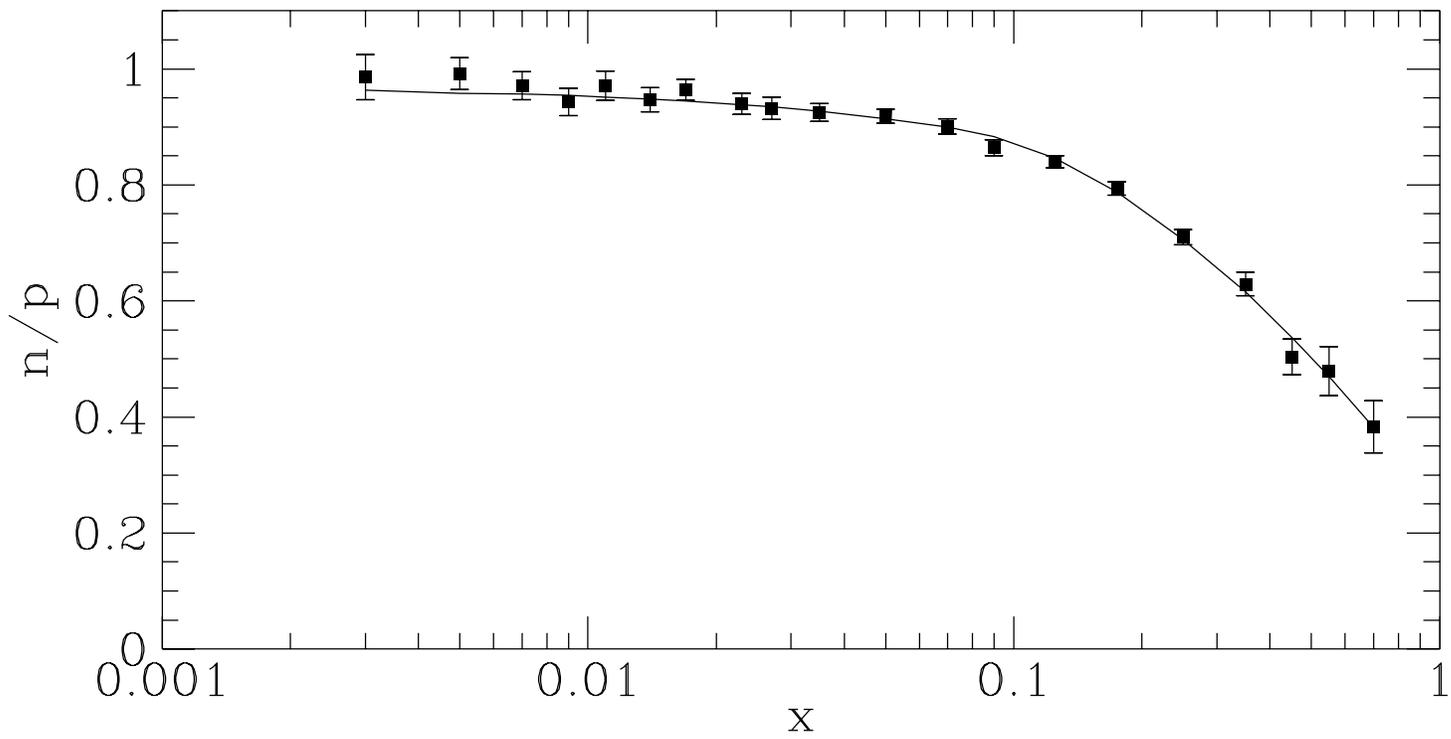
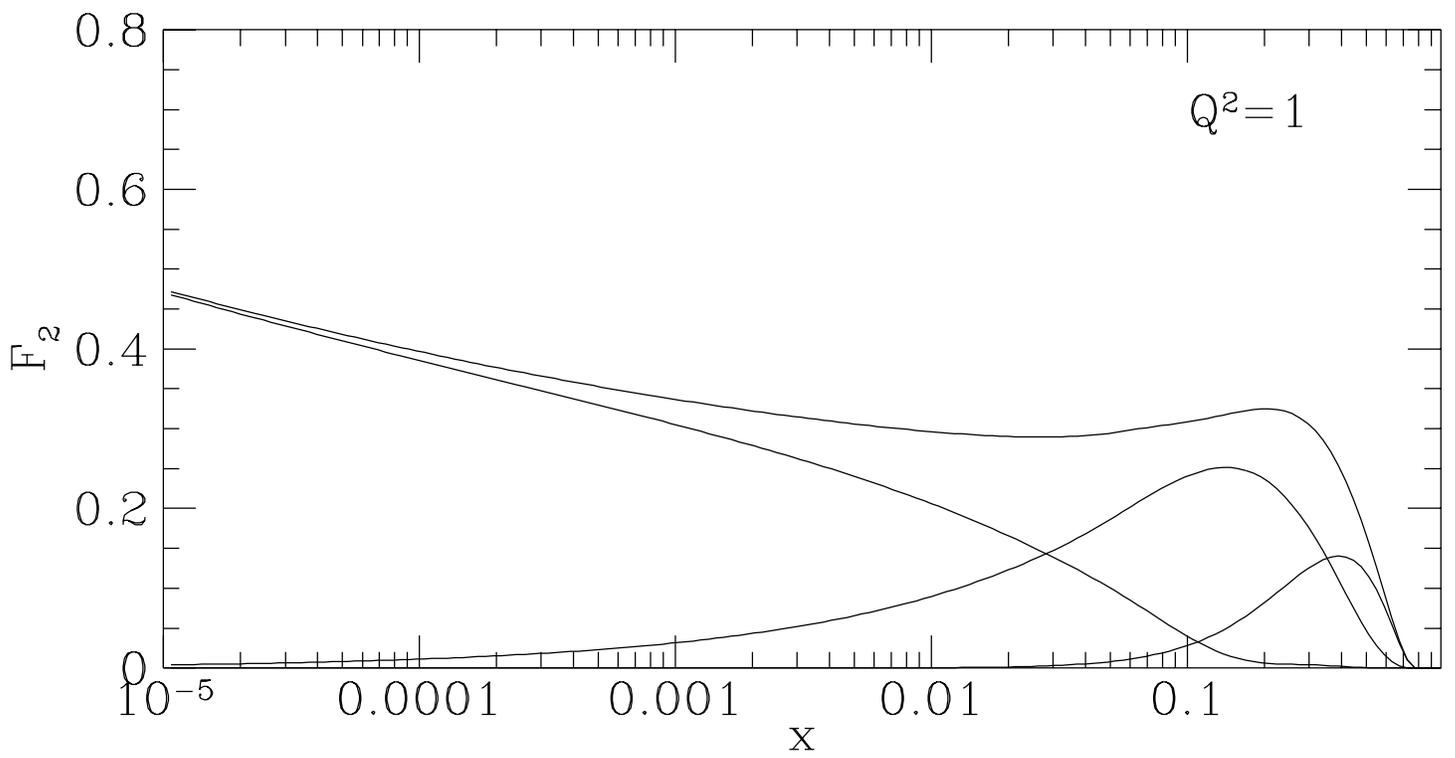
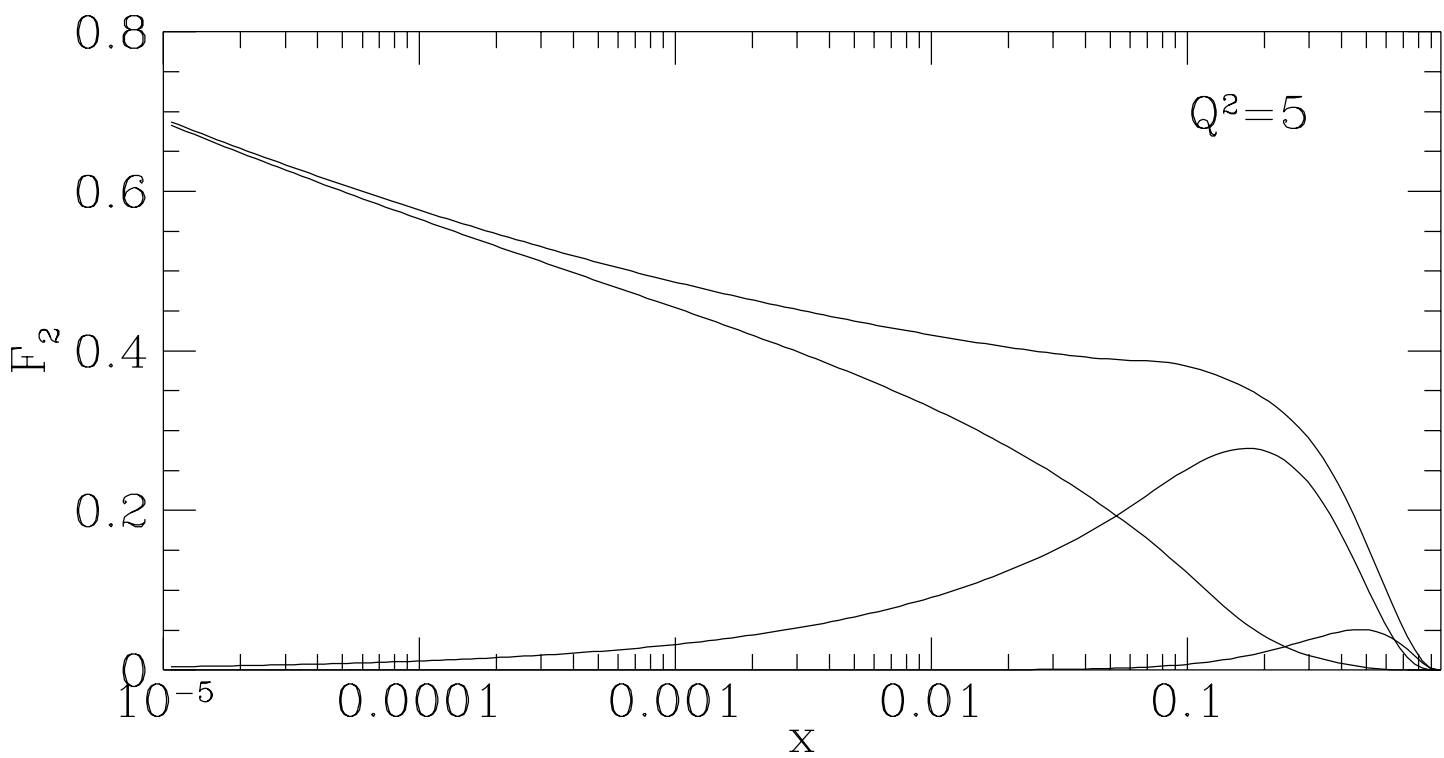
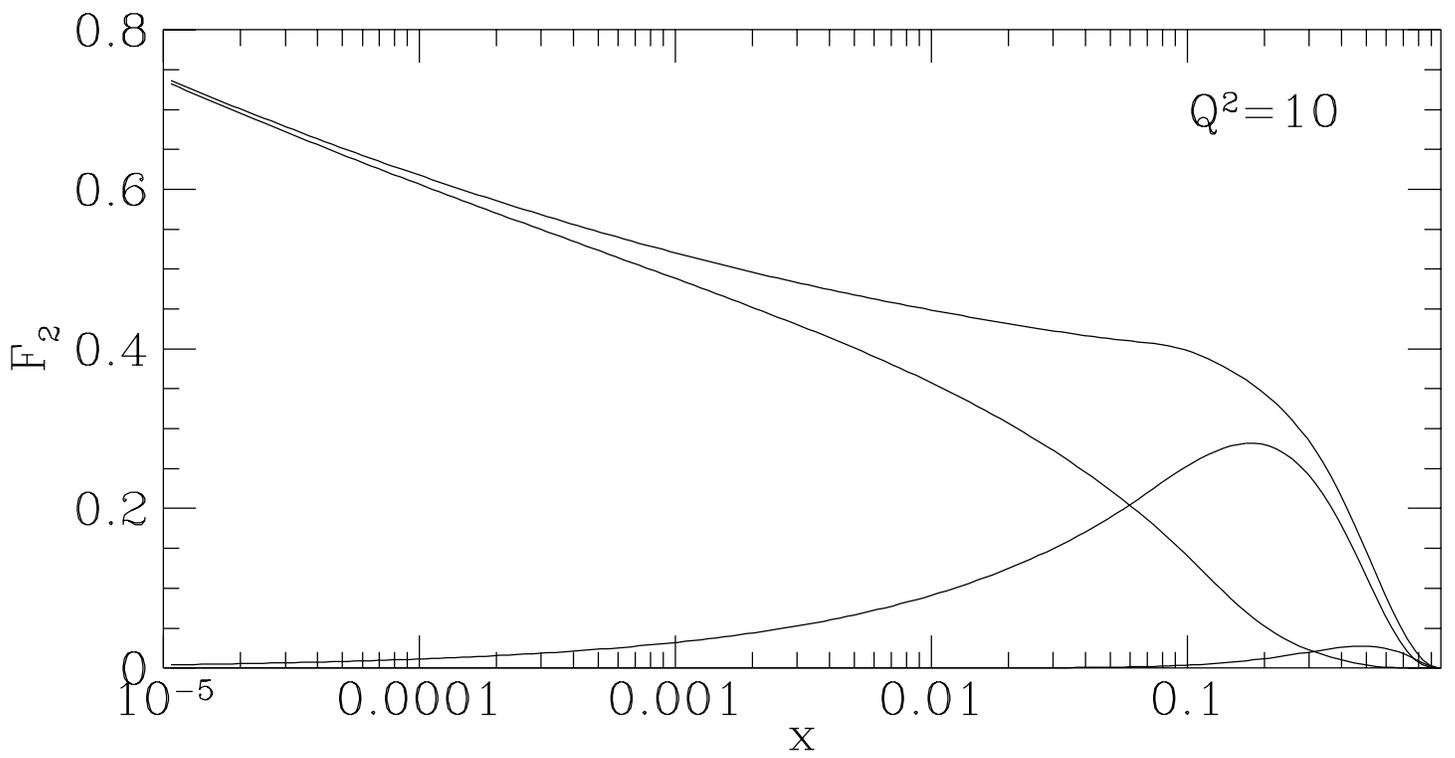
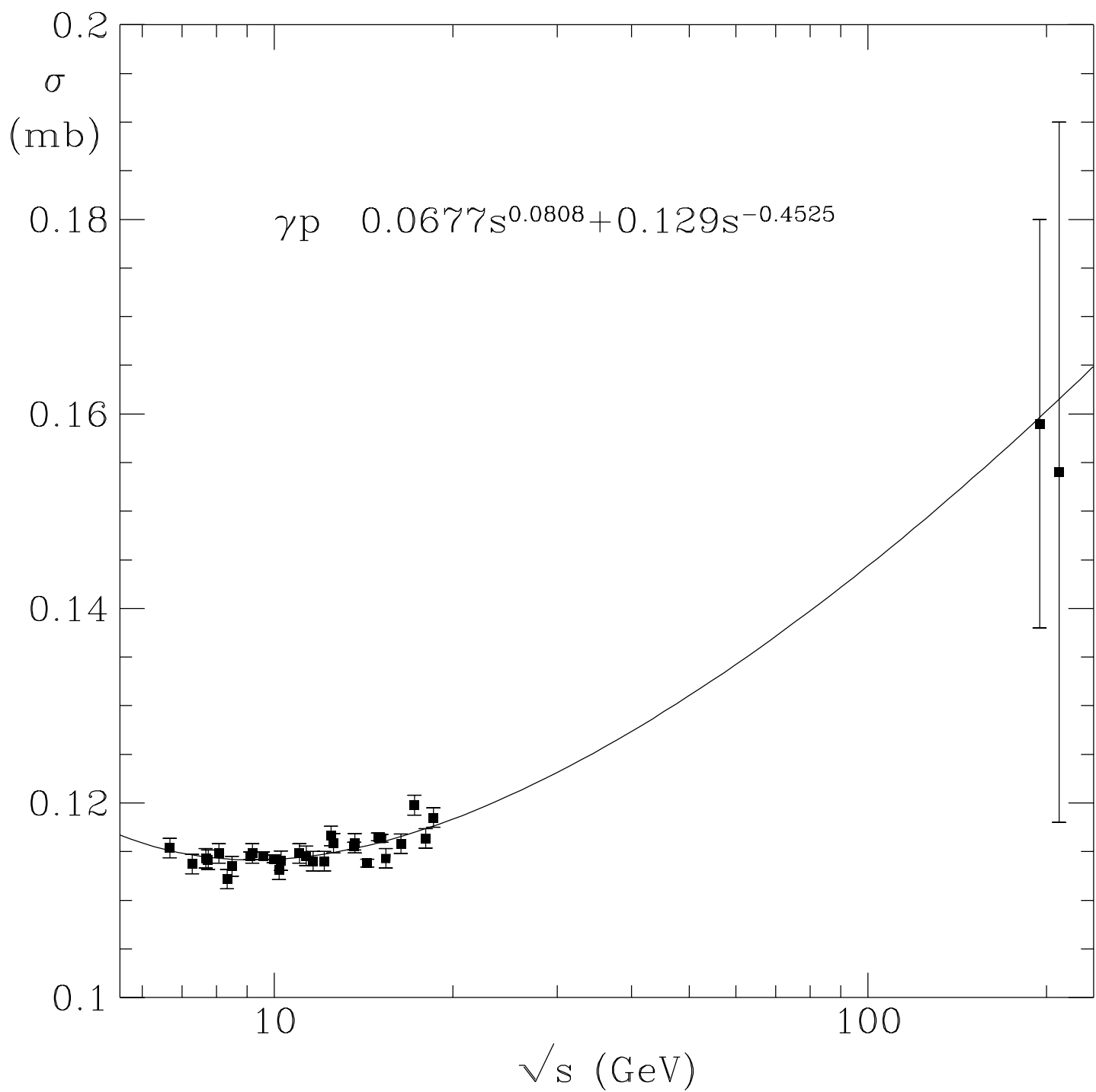
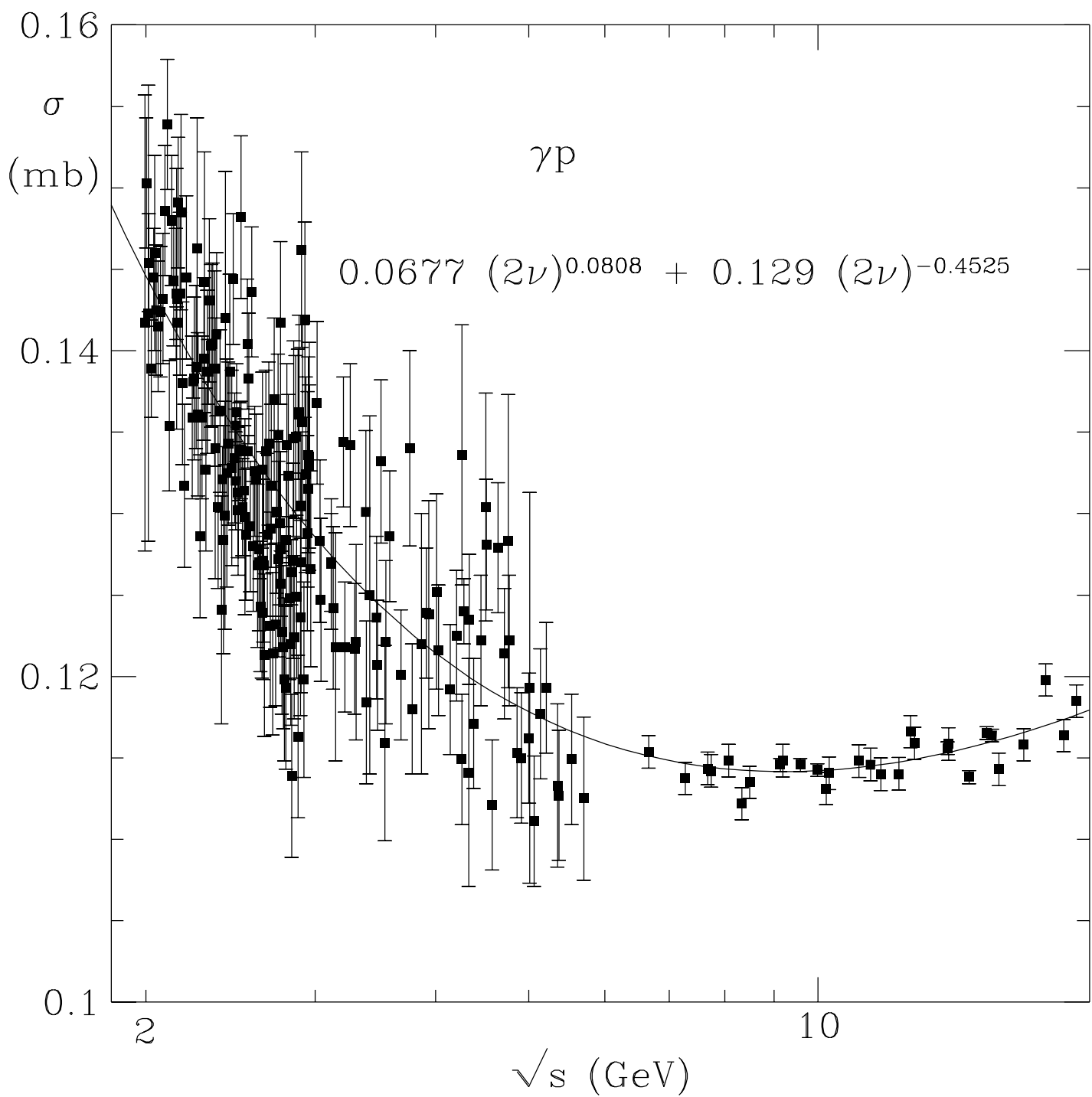
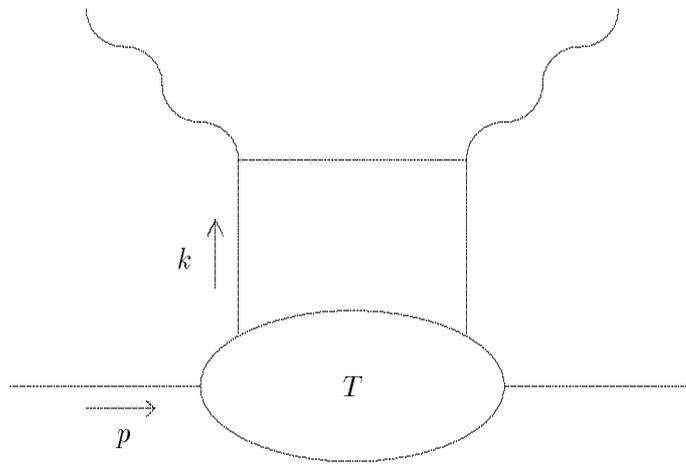
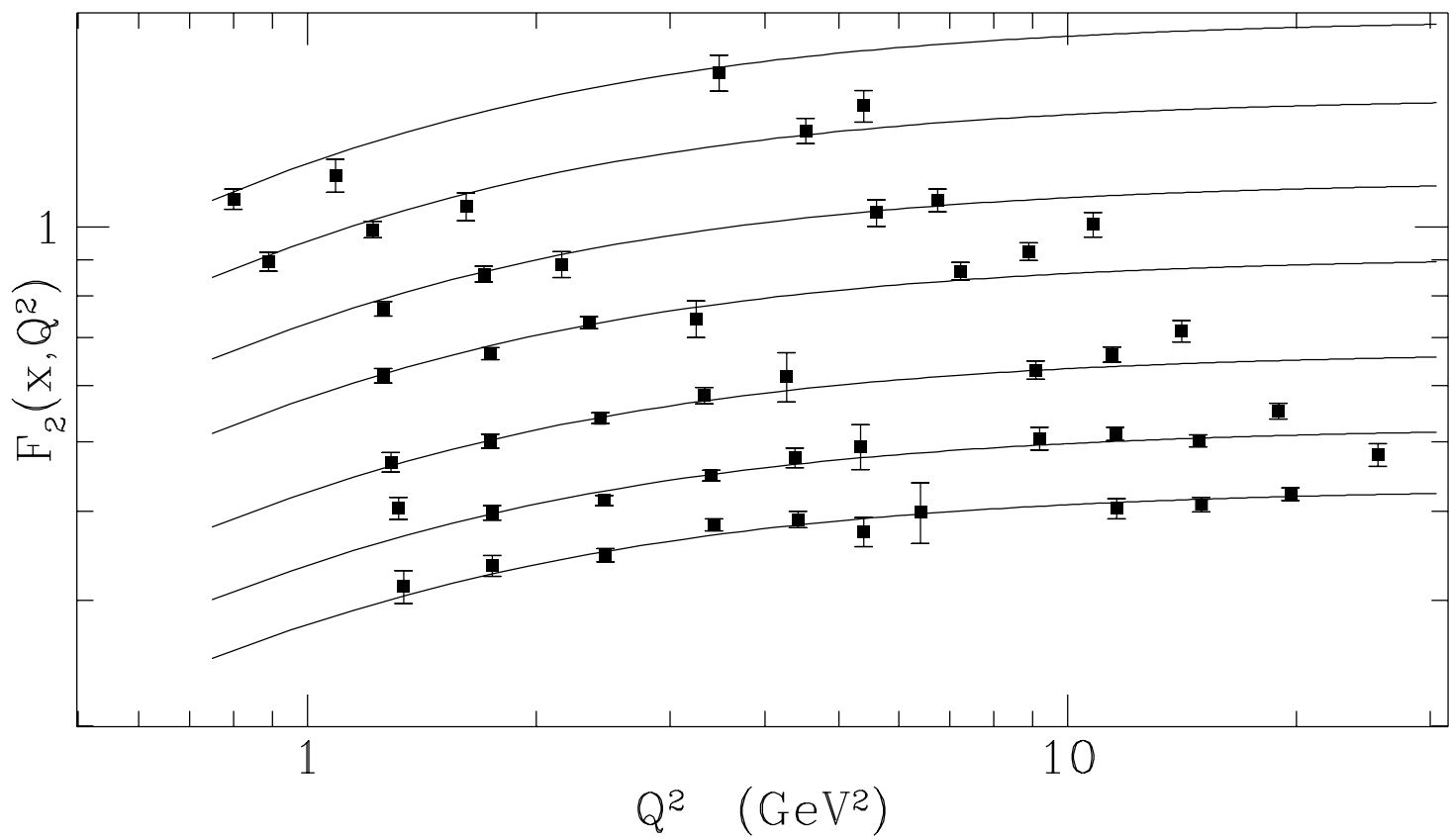
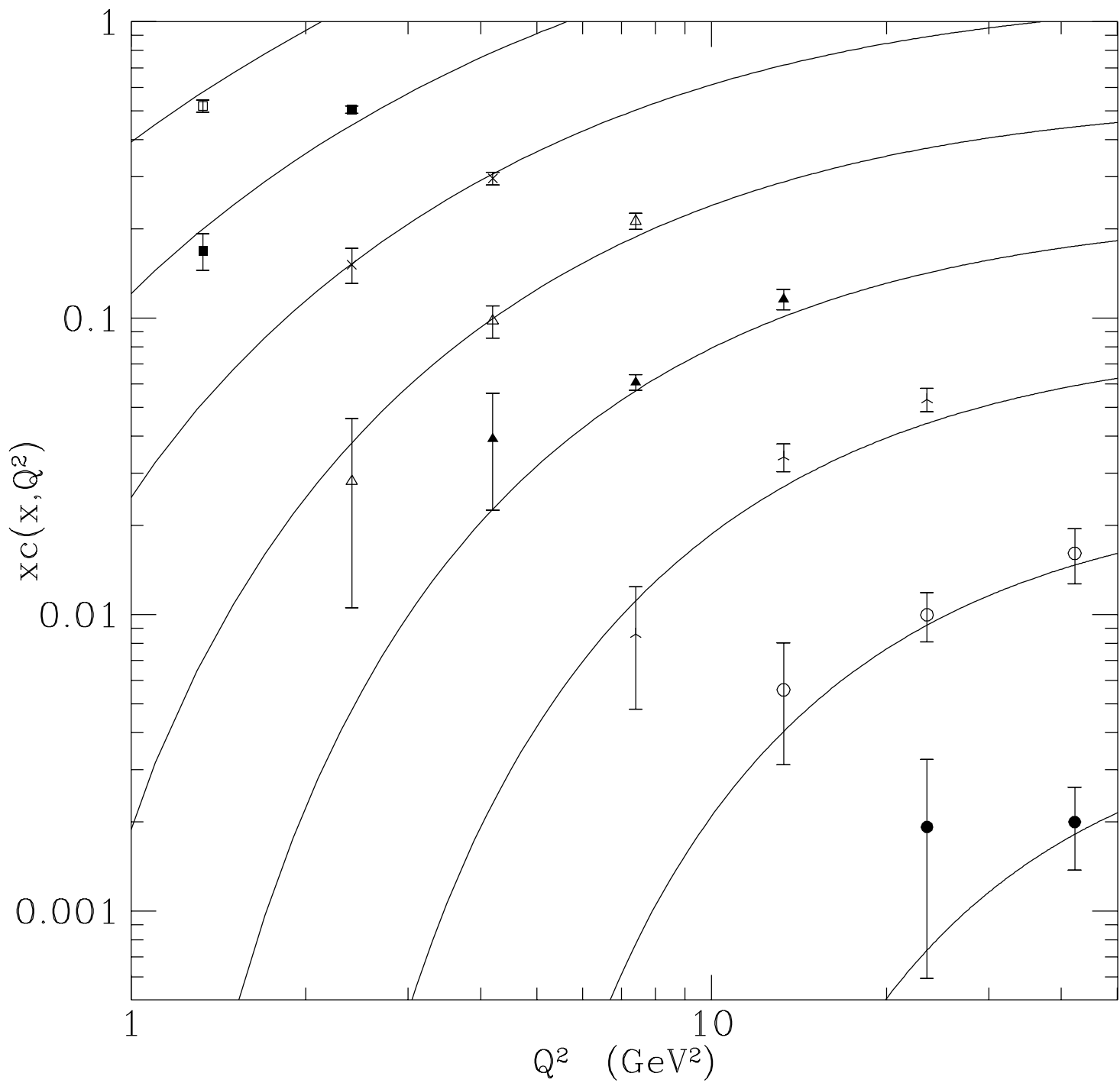
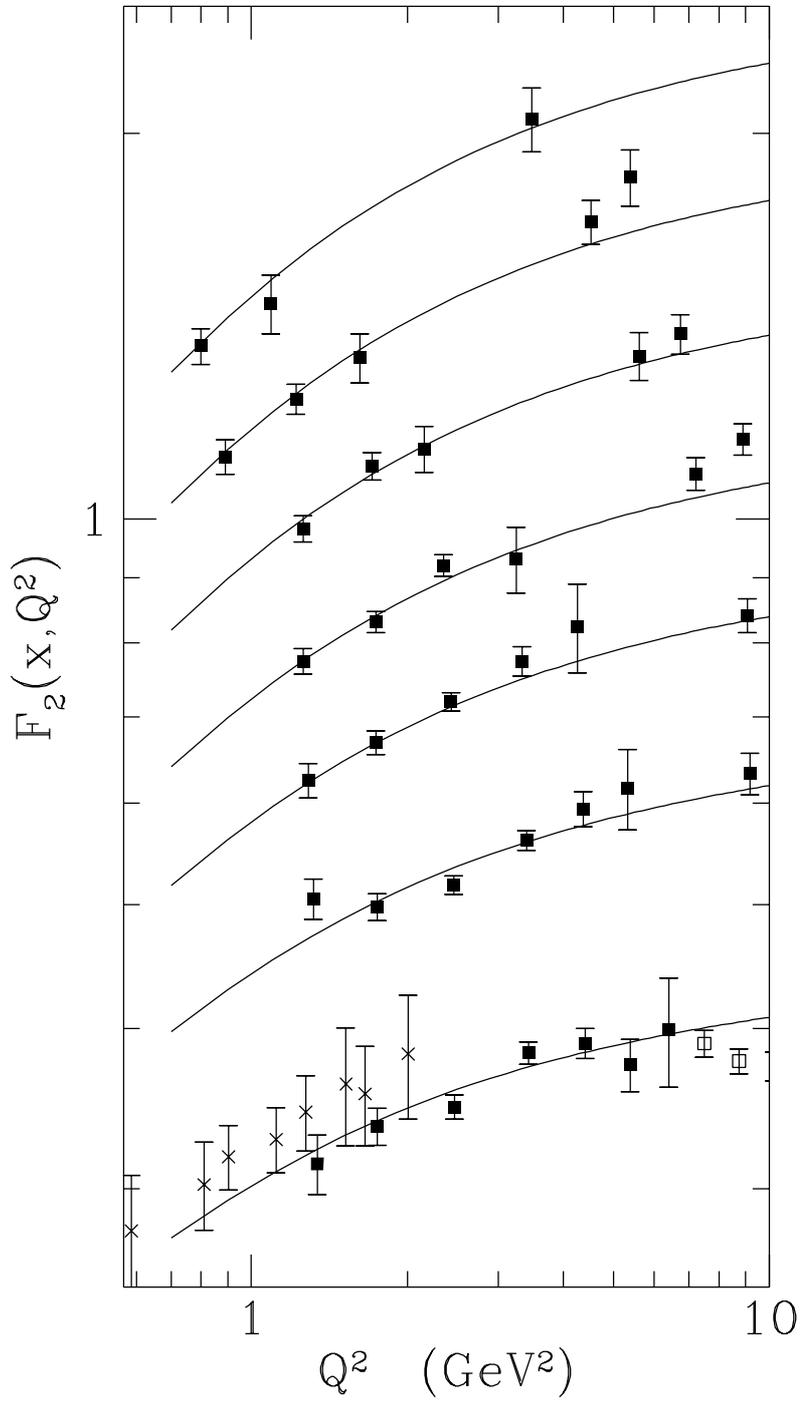
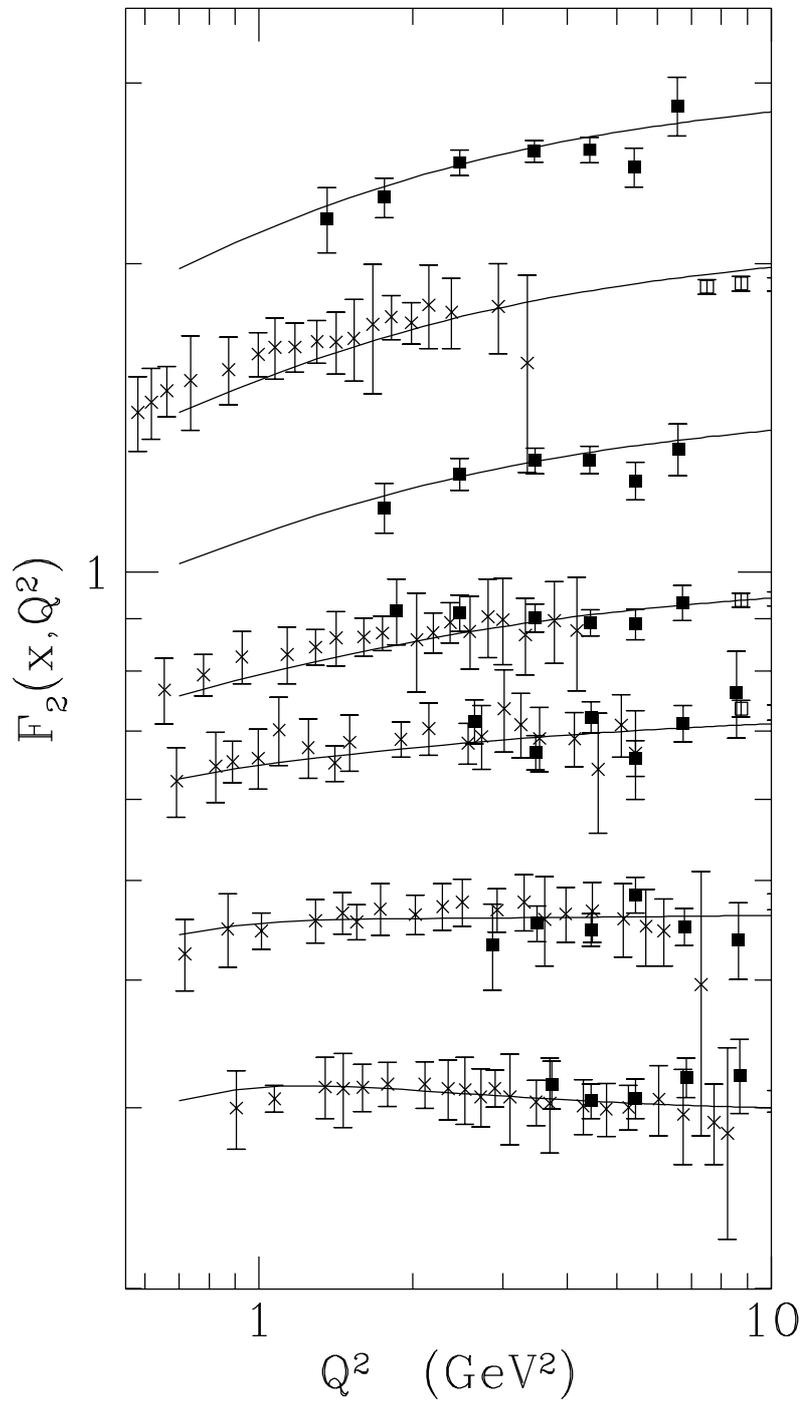
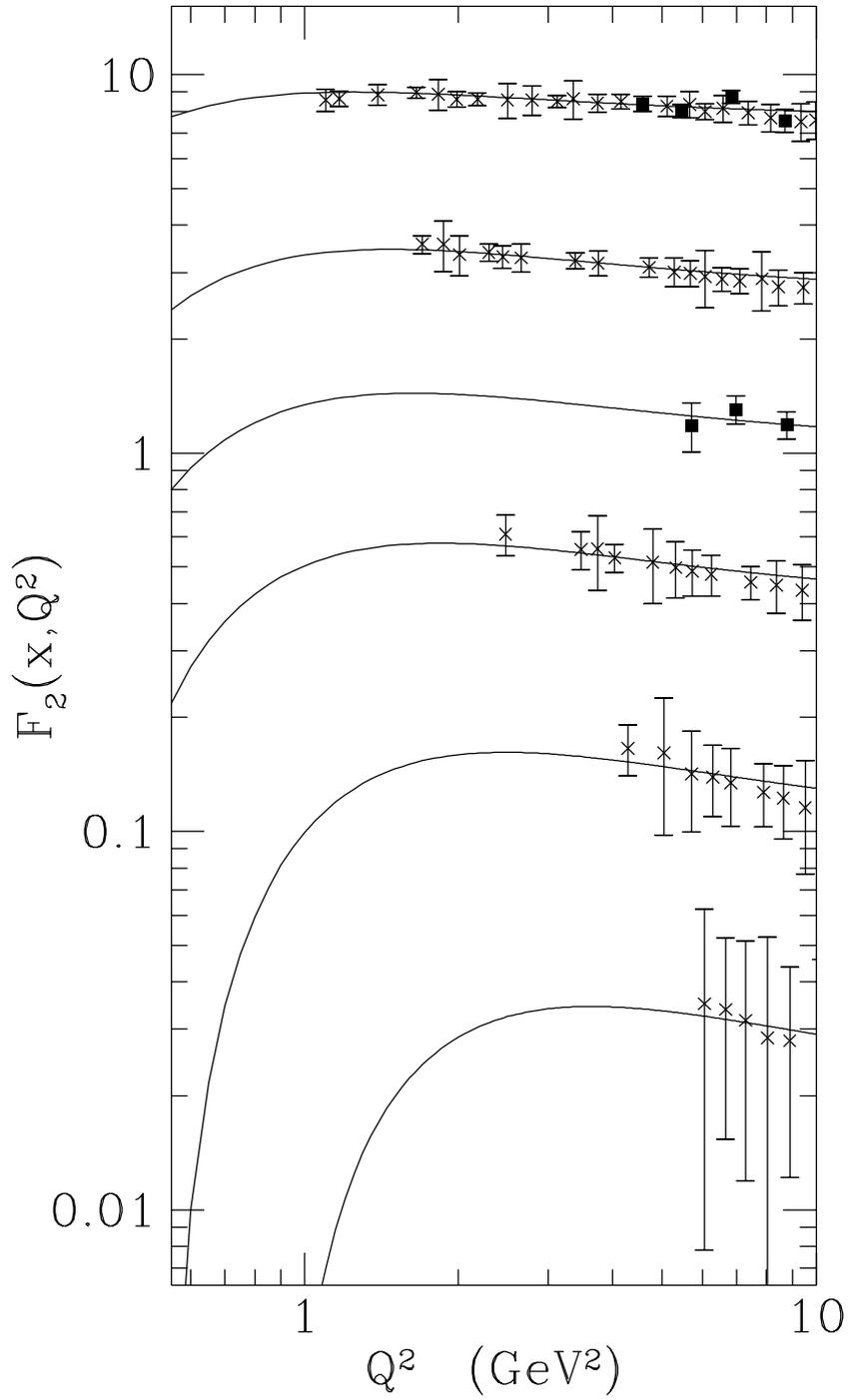
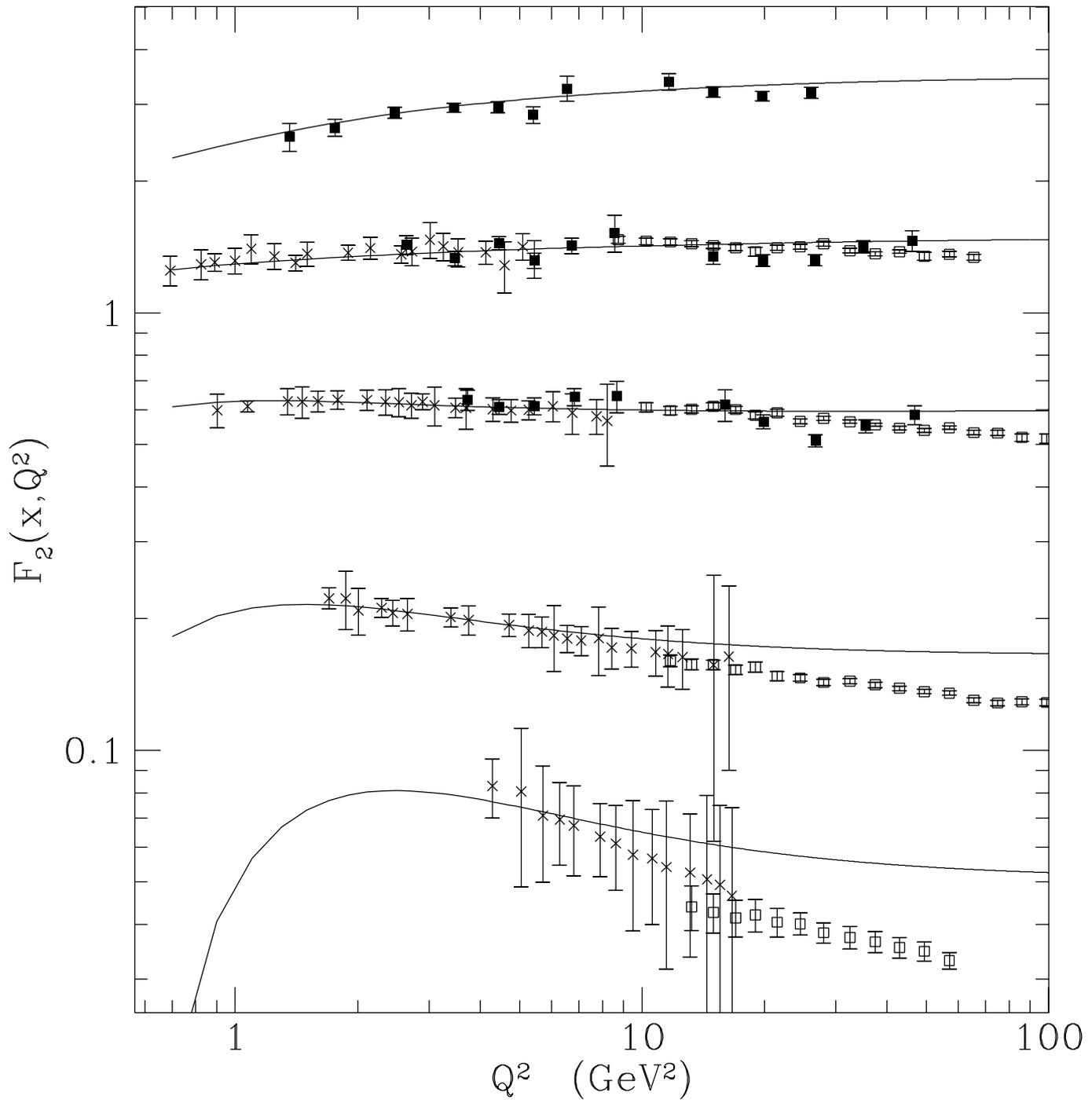
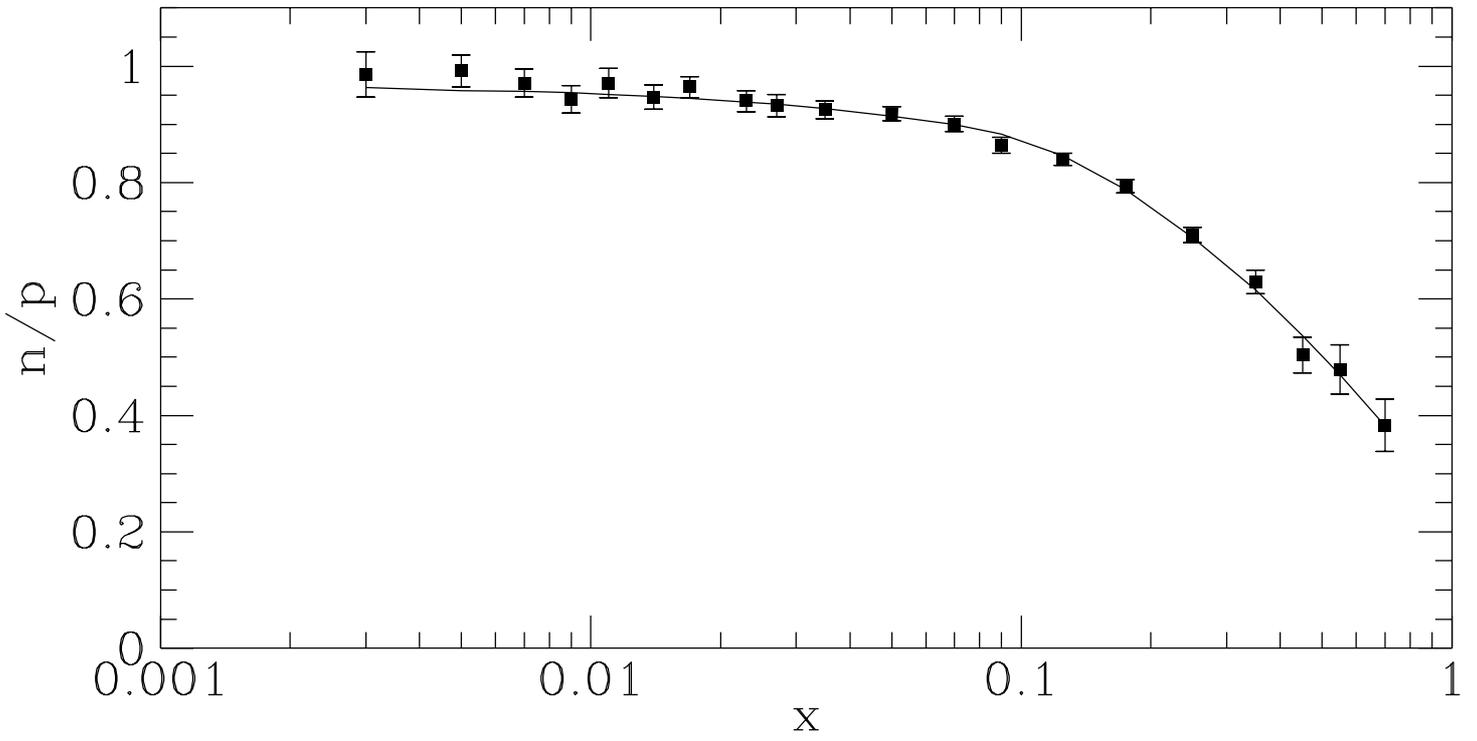
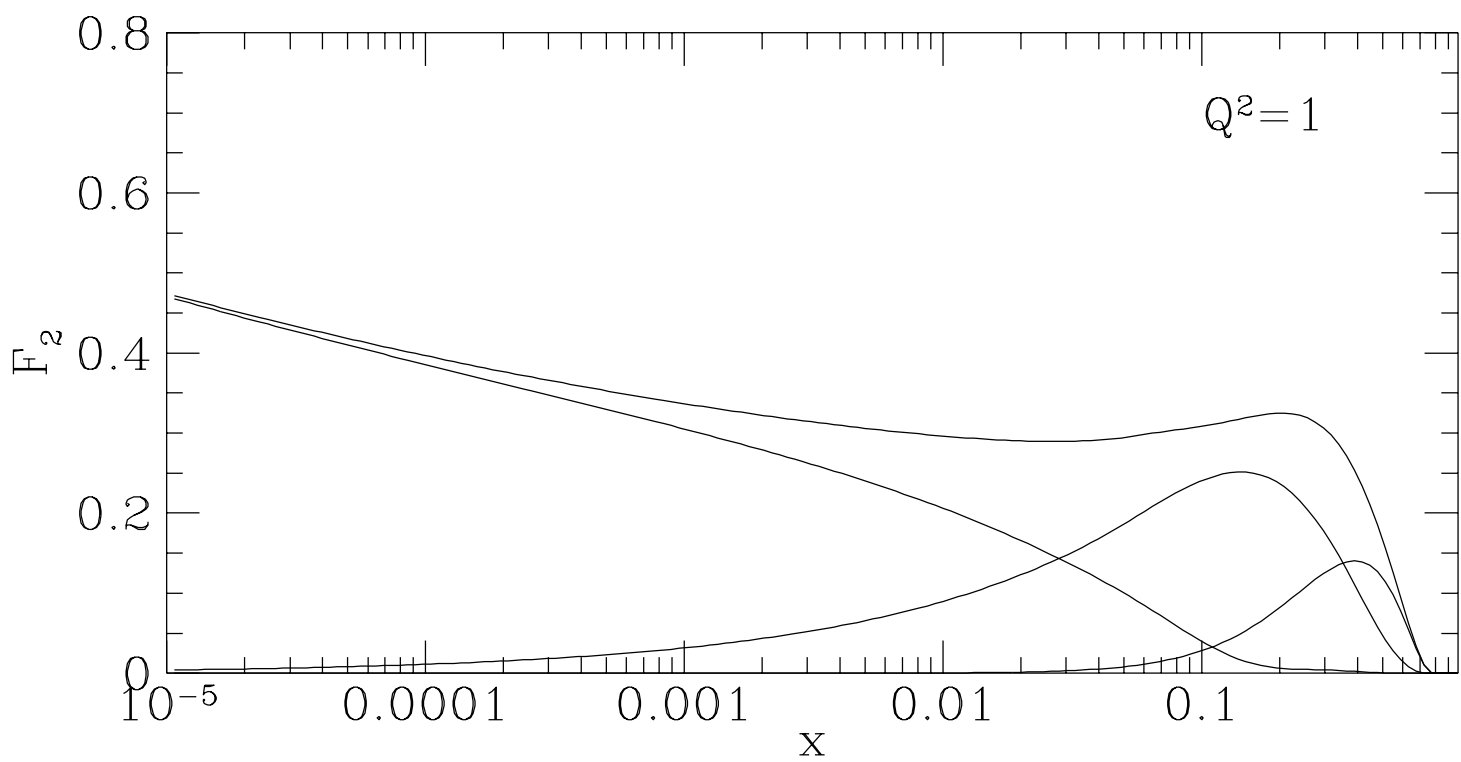
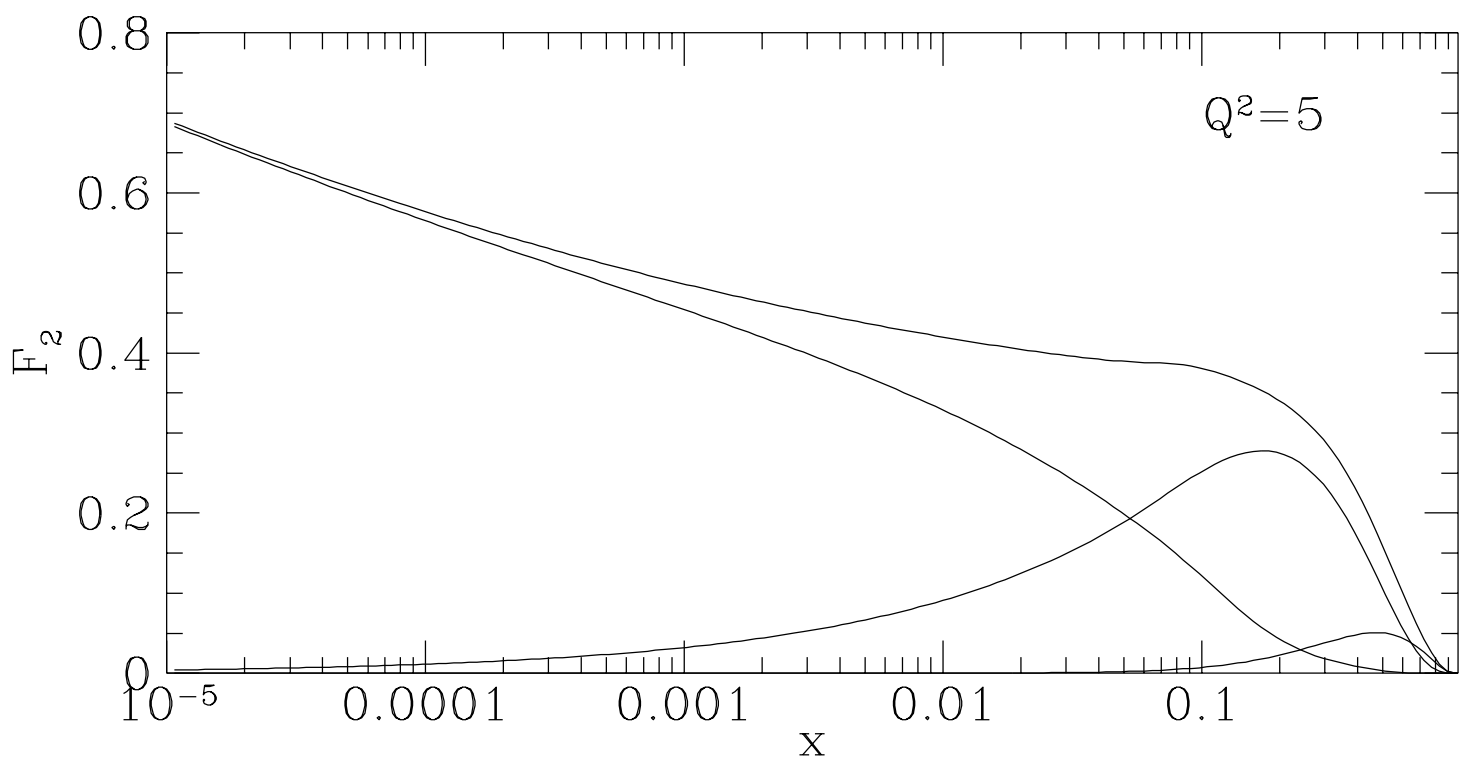
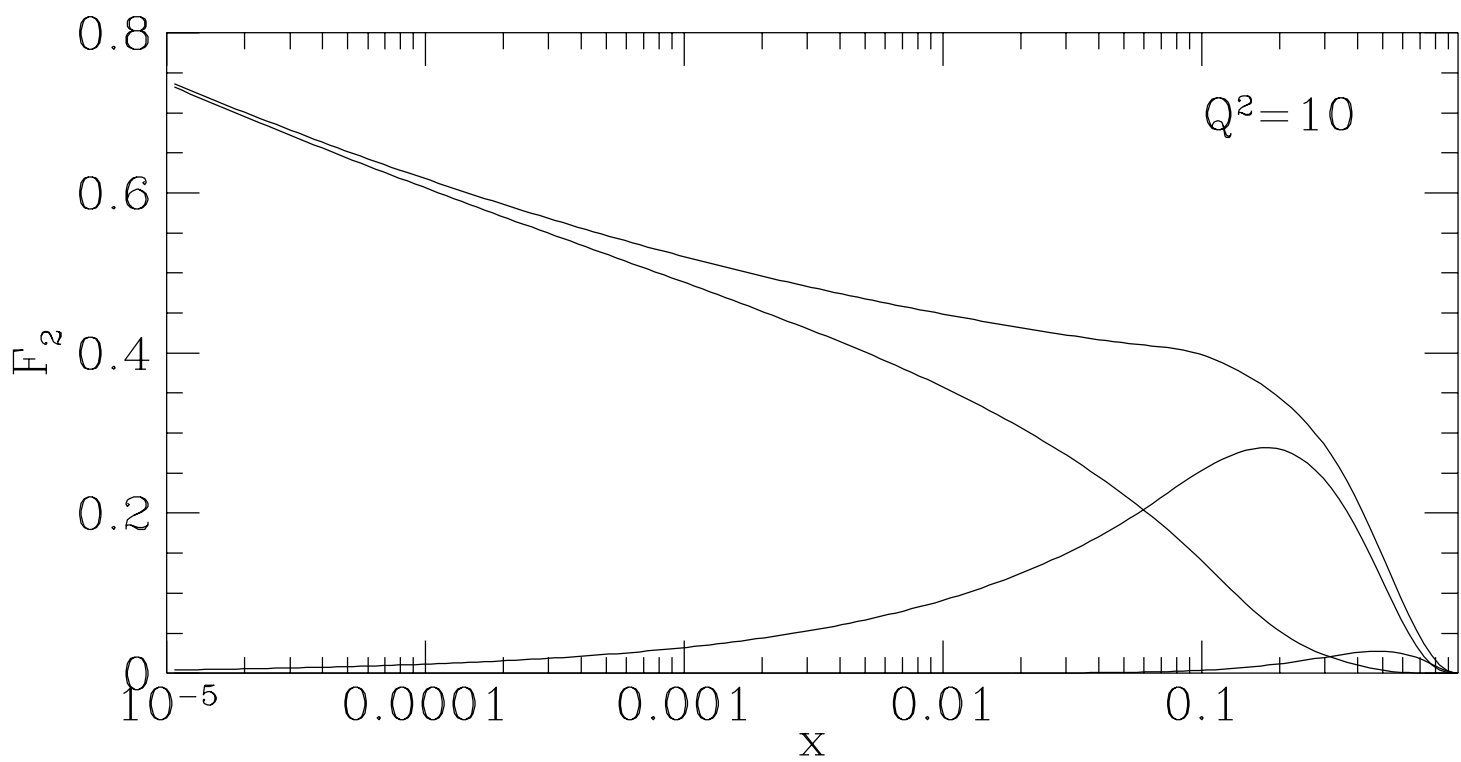
\bye